\documentclass{elsart}

\usepackage{natbib}

\usepackage{graphicx}
 
\begin{document}



\runauthor{Chengjin, Garrett, Nair, Porcas, Patnaik}


\begin{frontmatter} 
\title{Changes in the angular separation of the lensed images 
PKS 1830-211~NE \& SW}

\author[BAO,JIVE,MPIfR]{Jin Chengjin}
\author[JIVE]{M.A. Garrett}
\author[RAMAN]{S Nair}
\author[MPIfR]{R.W. Porcas}
\author[MPIfR]{A.R. Patnaik}

\address[BAO]{Beijing Astronomical Observatory, Chinese Academy of 
Sciences, A20 Datun Road, Chaoyang District, Beijing 100012, China}
\address[JIVE]{Joint Institute for VLBI in Europe, Postbus 2, 
7990~AA Dwingeloo, The~Netherlands} 
\address[MPIfR]{Max-Planck-Institut f\"ur Radioastronomie, 
Auf dem H\"ugel~69,D-53121~Bonn,Germany}
\address[RAMAN]{Raman Research Institute, C.V. Raman Avenue,
Bangalore 560080, India}
 

\begin{abstract}

  We present 8 epochs of 7\,mm dual-polarization VLBA observations of
  the gravitational lens system PKS~1830-211 made over the course of 14
  weeks. 
  Clear changes in the relative positions of the cores
  of up to $80\,\mu$arcseconds ($\mu$as) were observed between epochs (each
  separated by $\approx 2$~weeks). A comparison with previous 7~mm VLBA
  maps shows that the separation of the cores has changed by
  almost $280\,\mu$as over 12 months. This leads us to conclude that
  changes in the brightness distribution of the mm-VLBI core of the
  background source must be occurring rapidly.  This is
  the first clear observation of significant radio source evolution in
  a gravitational lens system. It is also the first time that changes
  in source structure have been detected in a distant extra-galactic
  source on such short time-scales. This is partly accounted for by the
  magnification provided by the lens system.

\end{abstract} 


\begin{keyword}

cosmology: gravitational lensing: individual ( PKS 1830-211 )


\PACS 98.62.Sb \sep 98.54.Cm \sep 98.70.Dk

\end{keyword}

\end{frontmatter} 

\section{Introduction}
\label{intro} 

PKS~1830-211 is a very bright and highly variable radio source at cm-
and mm-wavelengths. As well as being a highly probable gravitational lens
system \citep{Rao88}, it is also identified by the EGRET
instrument as a strong source of gamma-rays \citep{Mattox97}. PKS~1830-211 is
one of only two known lens systems (the other is B0218+357) which are bright
and compact enough to be detected and imaged with mm-VLBI. 
All these observations suggest that the background source 
in this system is uncommon in many respects and can probably 
be best classified as a blazar. 

Relatively rapid
changes in the brightness distribution of the images had been reported
earlier \citep{Garrett97}, an effect that may be partly explained
by the magnification provided by the lens system. In the case
of PKS~1830-211, this magnification may be as large as 5--10
\citep{Kochanek92, Nair93}.
Recent spectroscopic observations in the near-IR using the NTT with 
clear detections of both  the H$_\alpha$ and H$_\beta$ emission lines (see
\citet{Lidman99}) have finally revealed the redshift of the source to be
$z_{s}=2.507$.
In this paper we present multi-epoch VLBA 7\,mm maps of both lensed
radio images, in both polarised and total intensity.

\section{Observations and Data Reduction} 
\label{features} 

We made eight epochs of 7\,mm, dual-polarisation VLBA observations of
PKS 1830-211 between 1997 January 19 and 1997 April 30.  Each epoch was
separated in time by about 14 days. The data were correlated at NRAO,
Socorro. For the sixth epoch (1997 April 03), the data quality was very poor 
partly due to bad weather at KP and weak fringes at BR. 
Since the SW and NE images are separated by about $1''$ on the
sky, wide-field techniques were used to make maps of both images
simultaneously from a single data-set (see \citet{Garrett99}). 
The polarisation data analysis followed \citet{Leppanen95}.

\section{Results \& Discussion} 
\label{results} 

\begin{figure}[htbp]
\centering
\includegraphics[scale=0.13,angle=0]{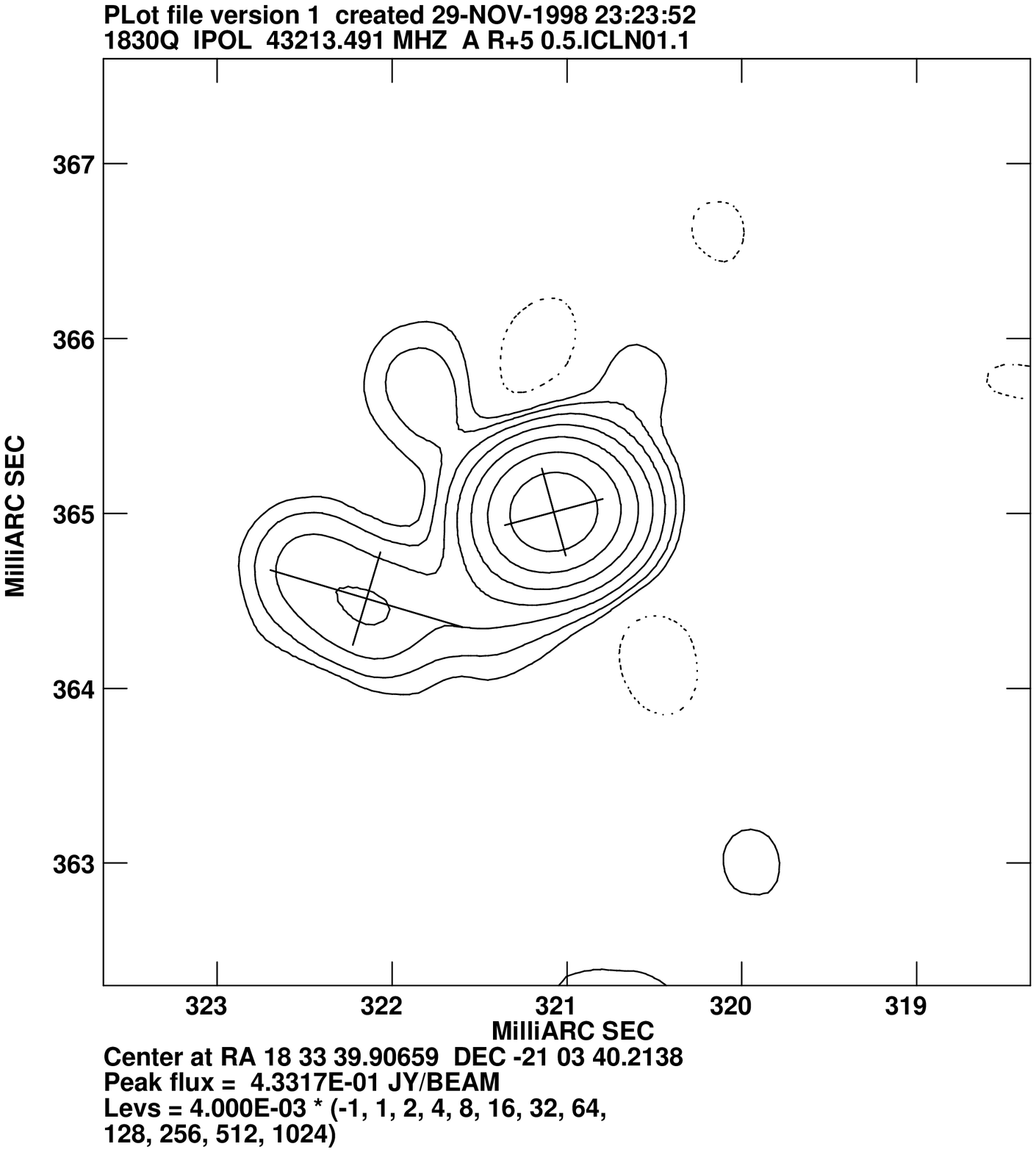}
\includegraphics[scale=0.13,angle=0]{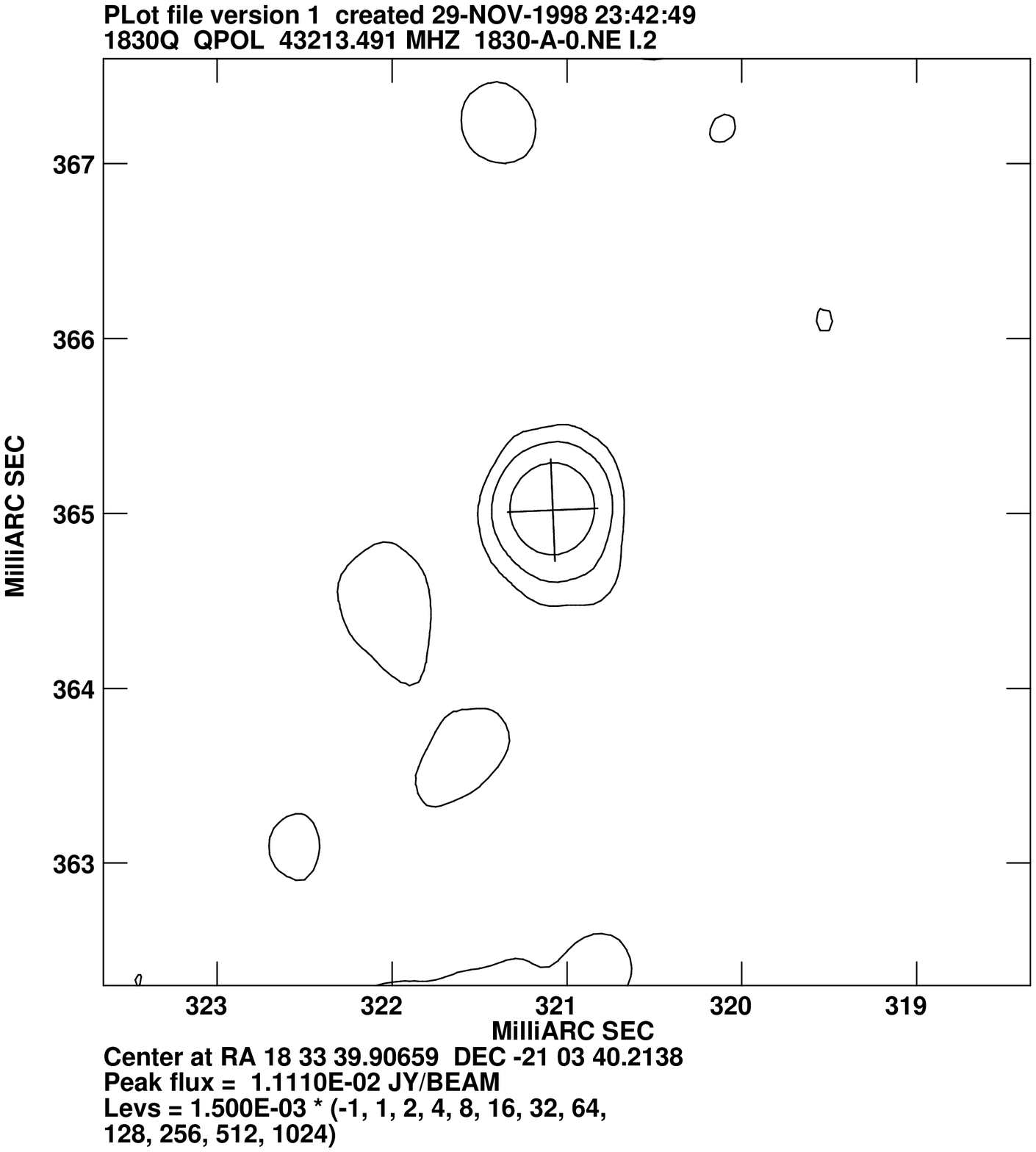}
\includegraphics[scale=0.13,angle=0]{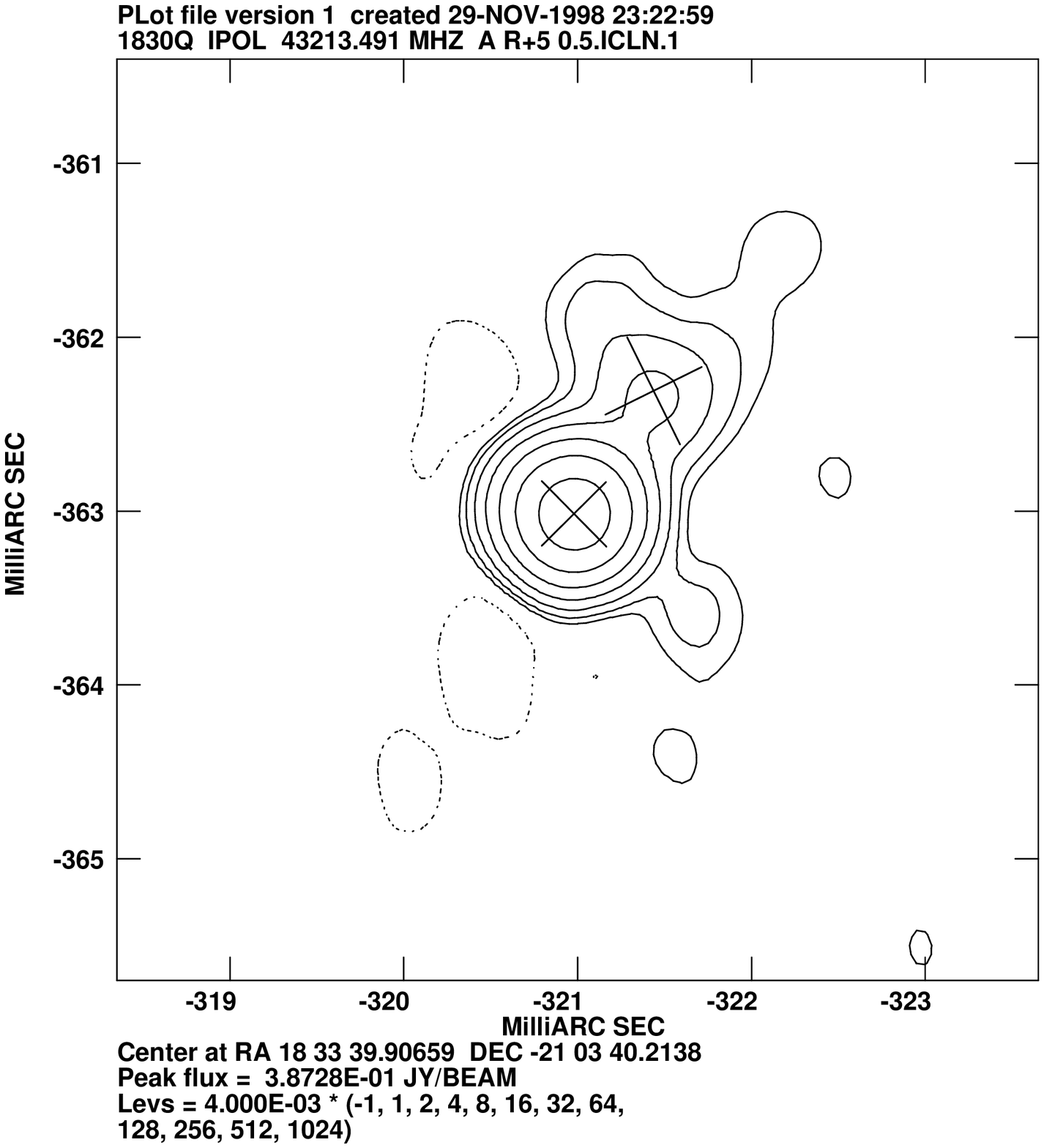}
\includegraphics[scale=0.13,angle=0]{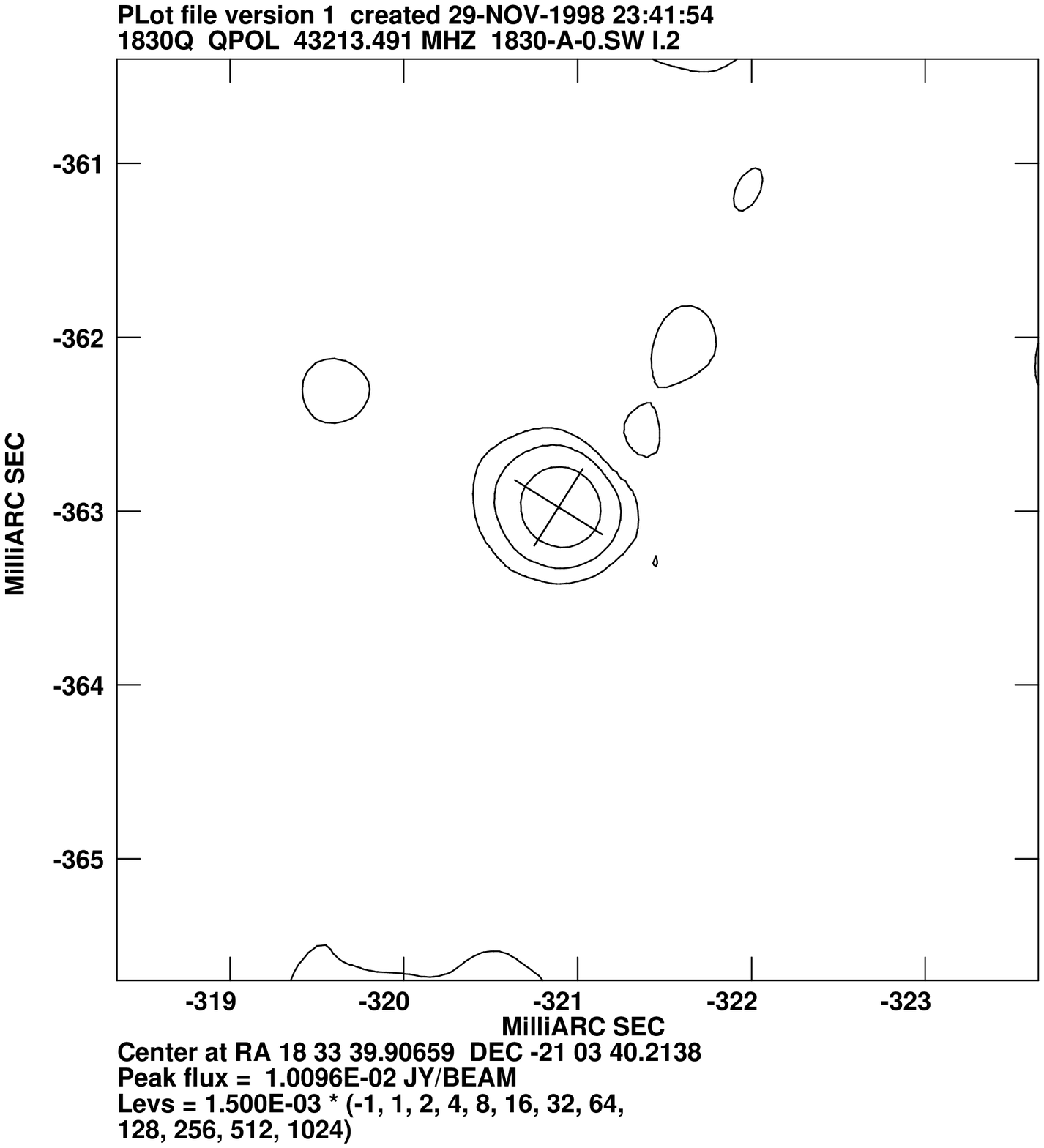}

\includegraphics[scale=0.13,angle=0]{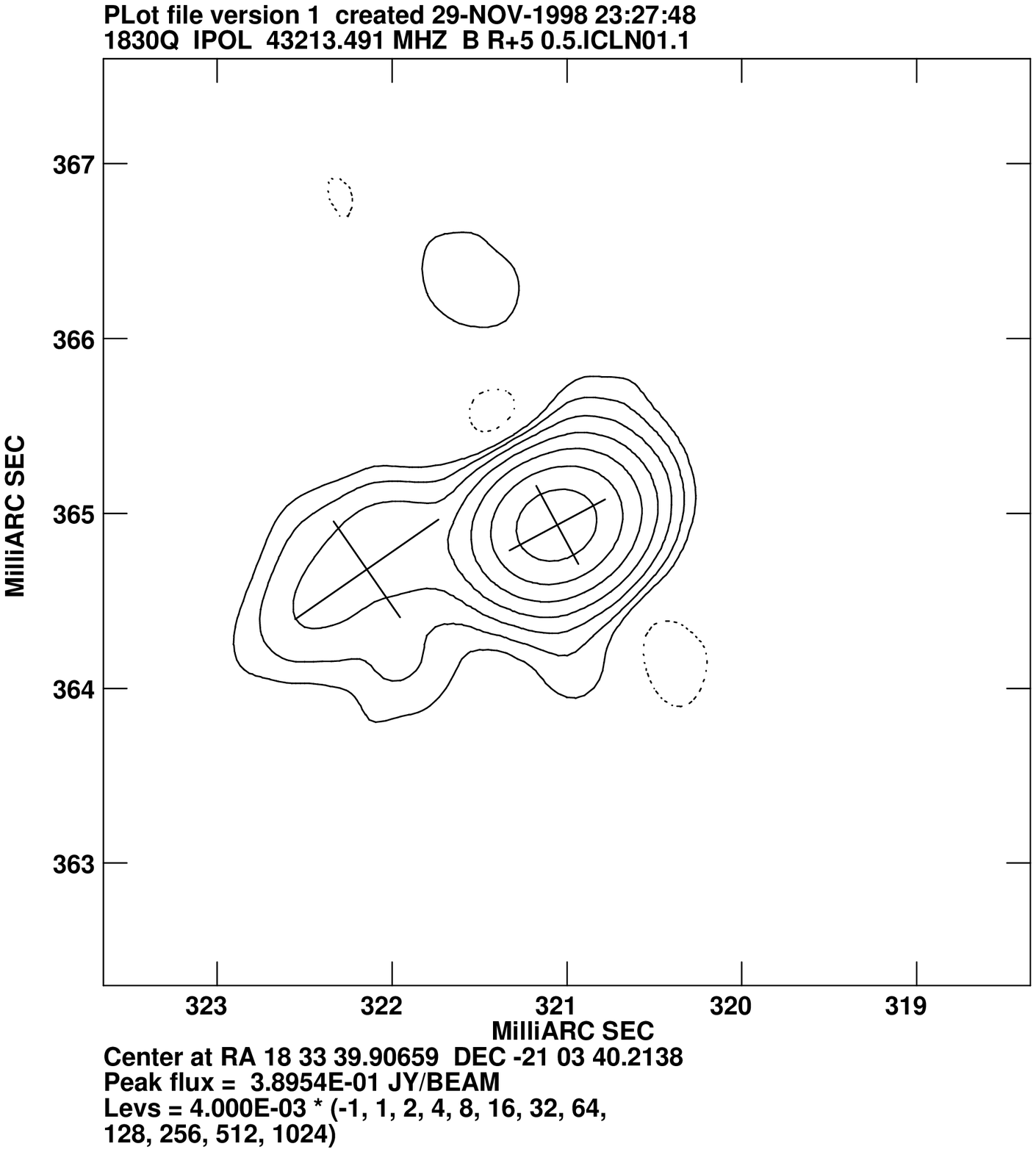}
\includegraphics[scale=0.13,angle=0]{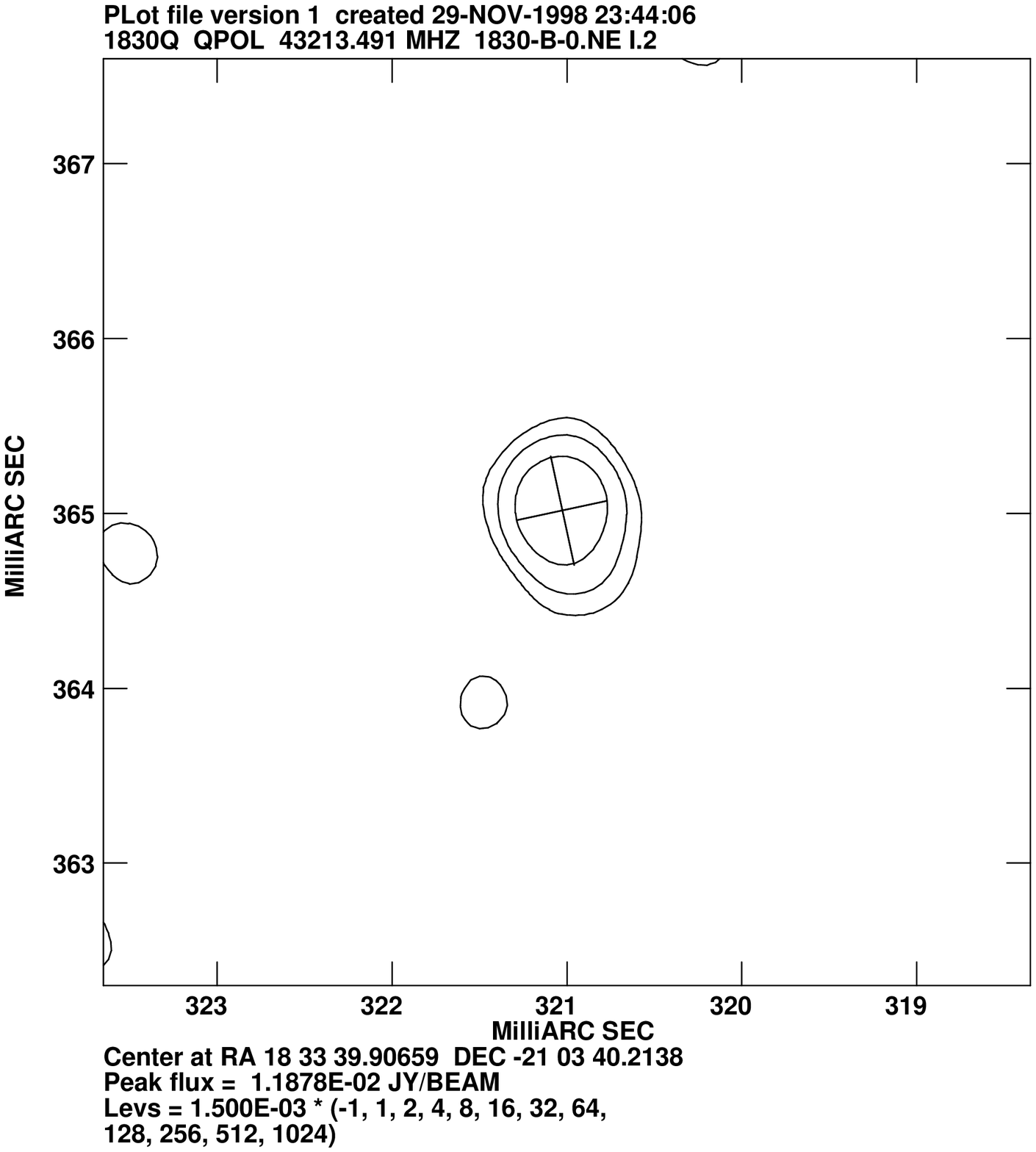}
\includegraphics[scale=0.13,angle=0]{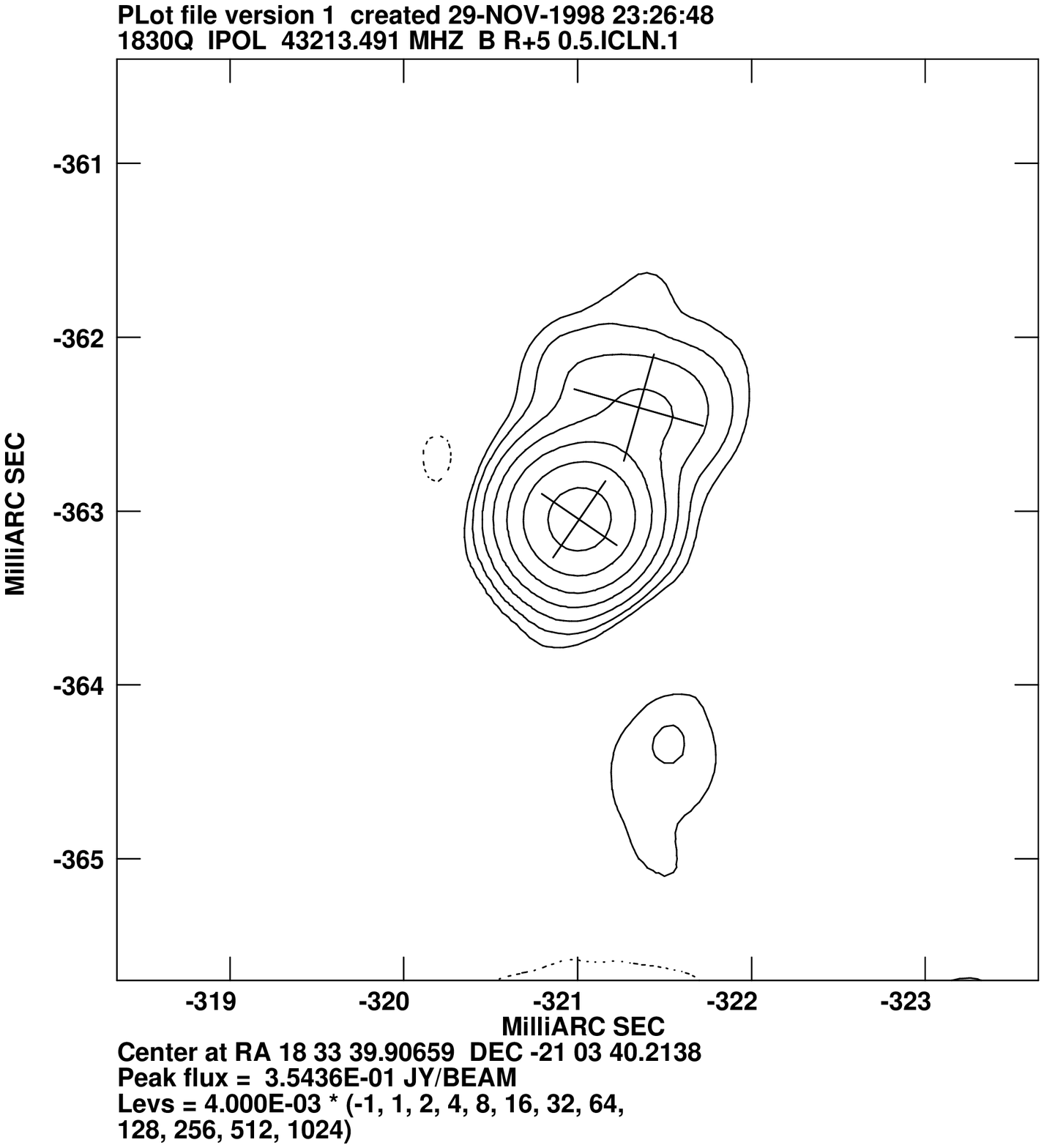}
\includegraphics[scale=0.13,angle=0]{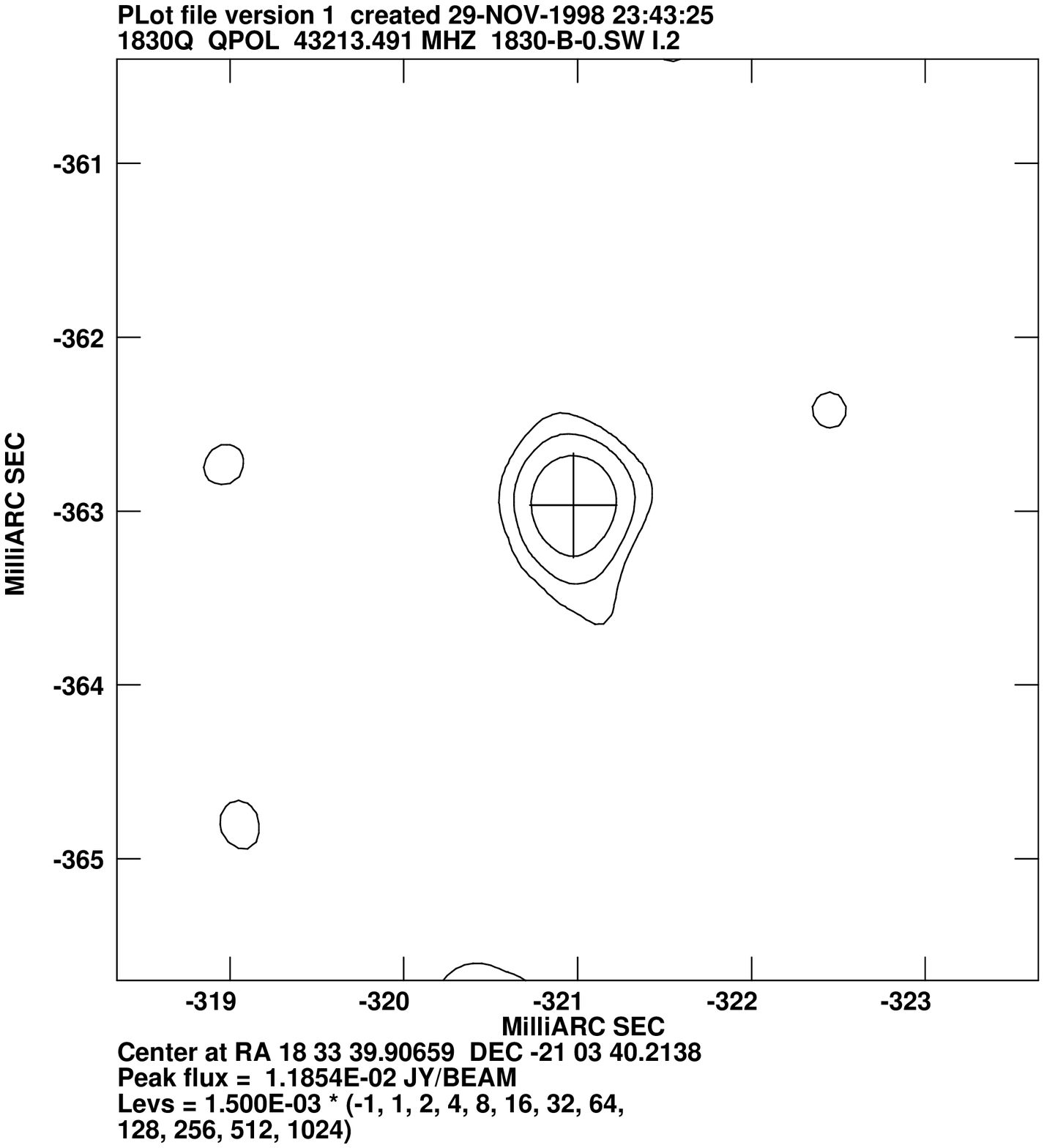}

\includegraphics[scale=0.13,angle=0]{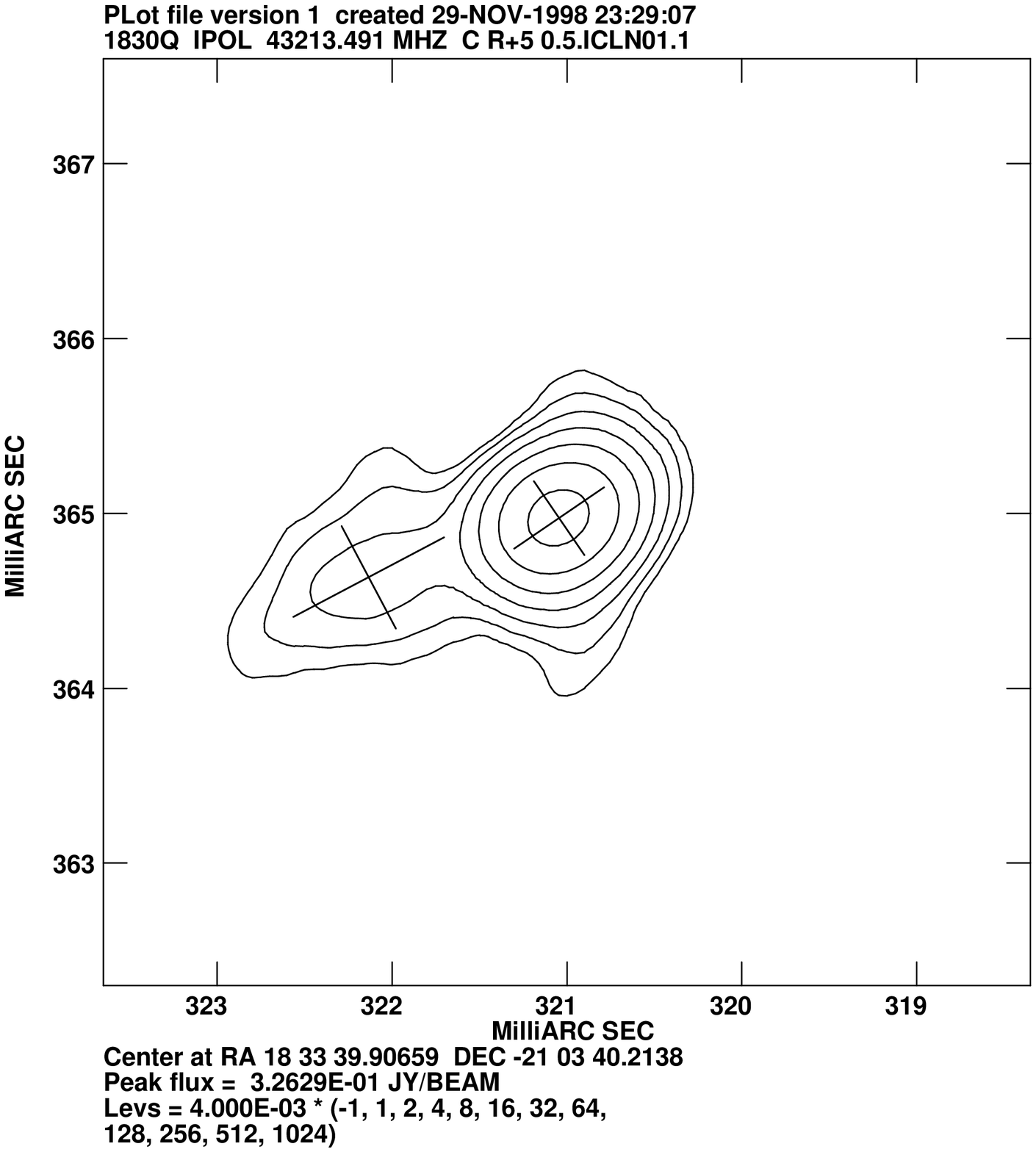}
\includegraphics[scale=0.13,angle=0]{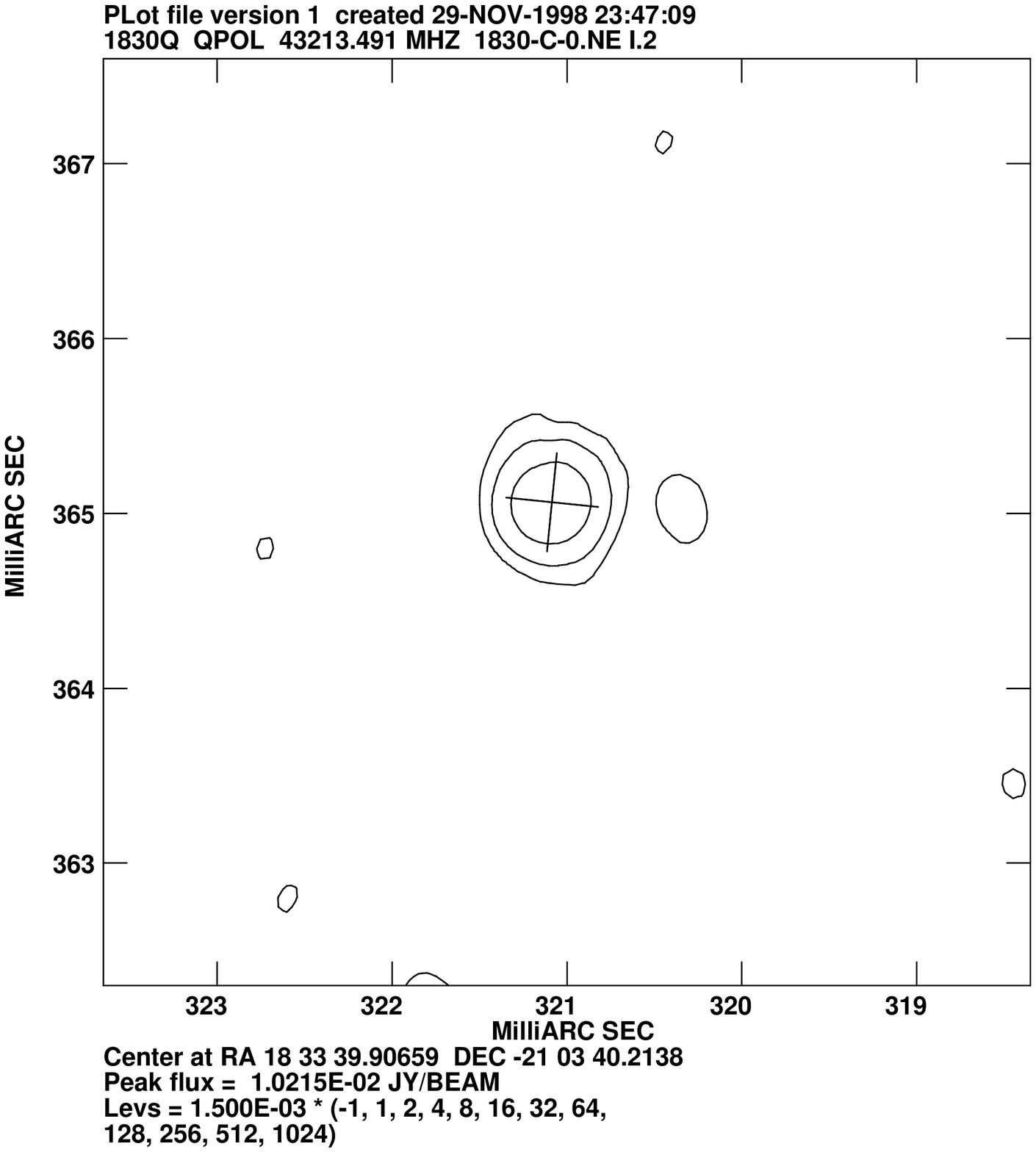}
\includegraphics[scale=0.13,angle=0]{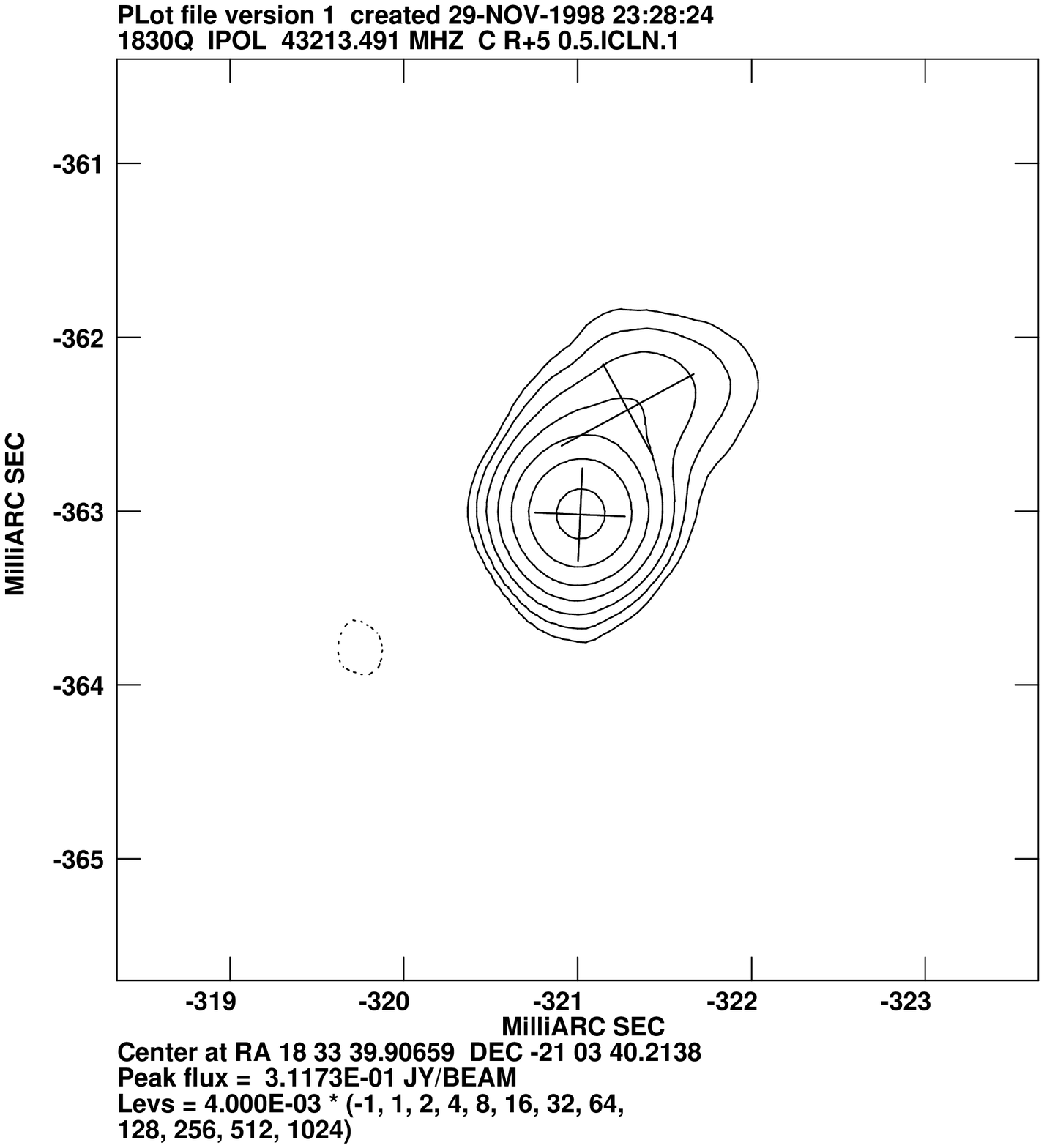}
\includegraphics[scale=0.13,angle=0]{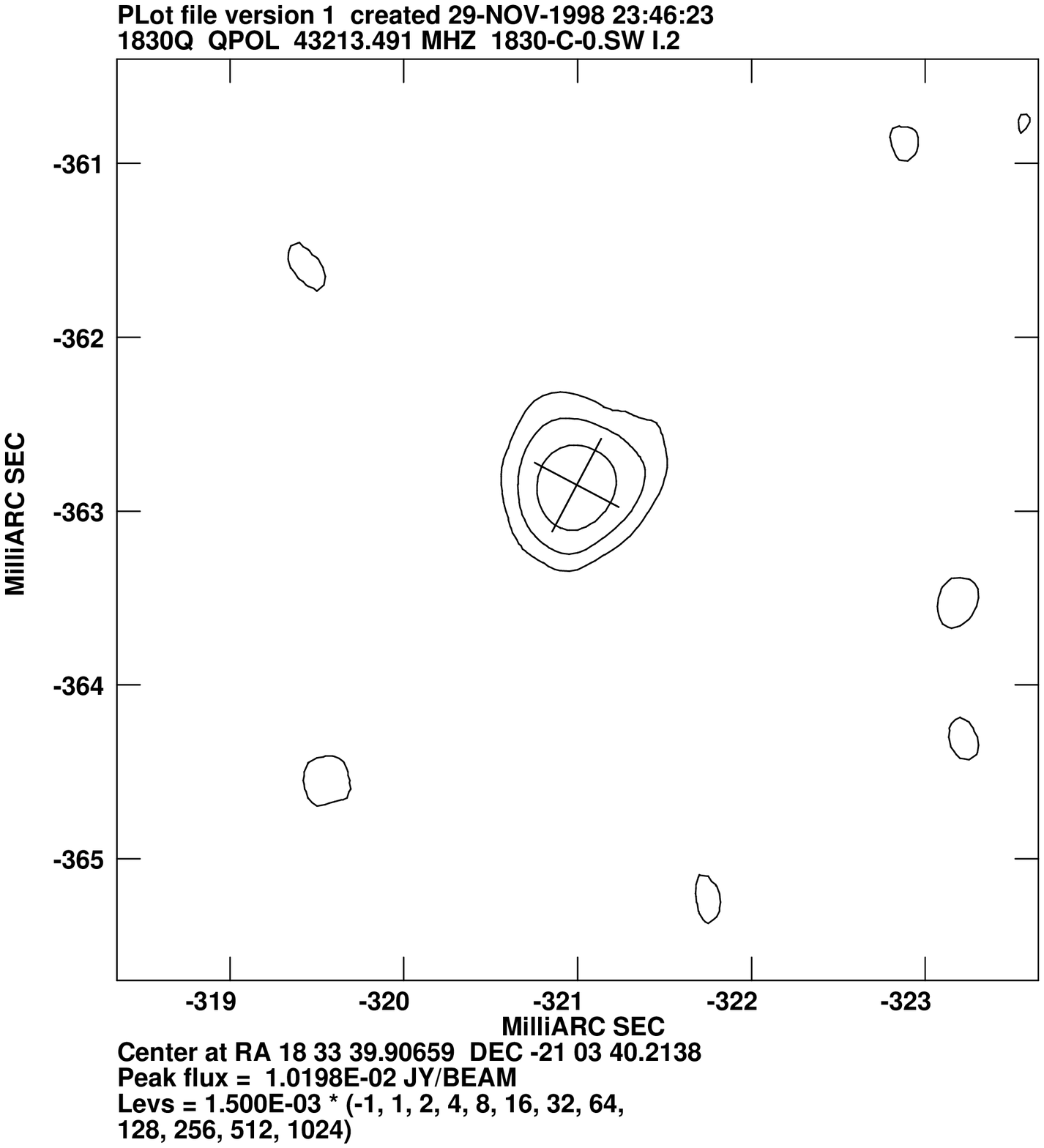}

\includegraphics[scale=0.13,angle=0]{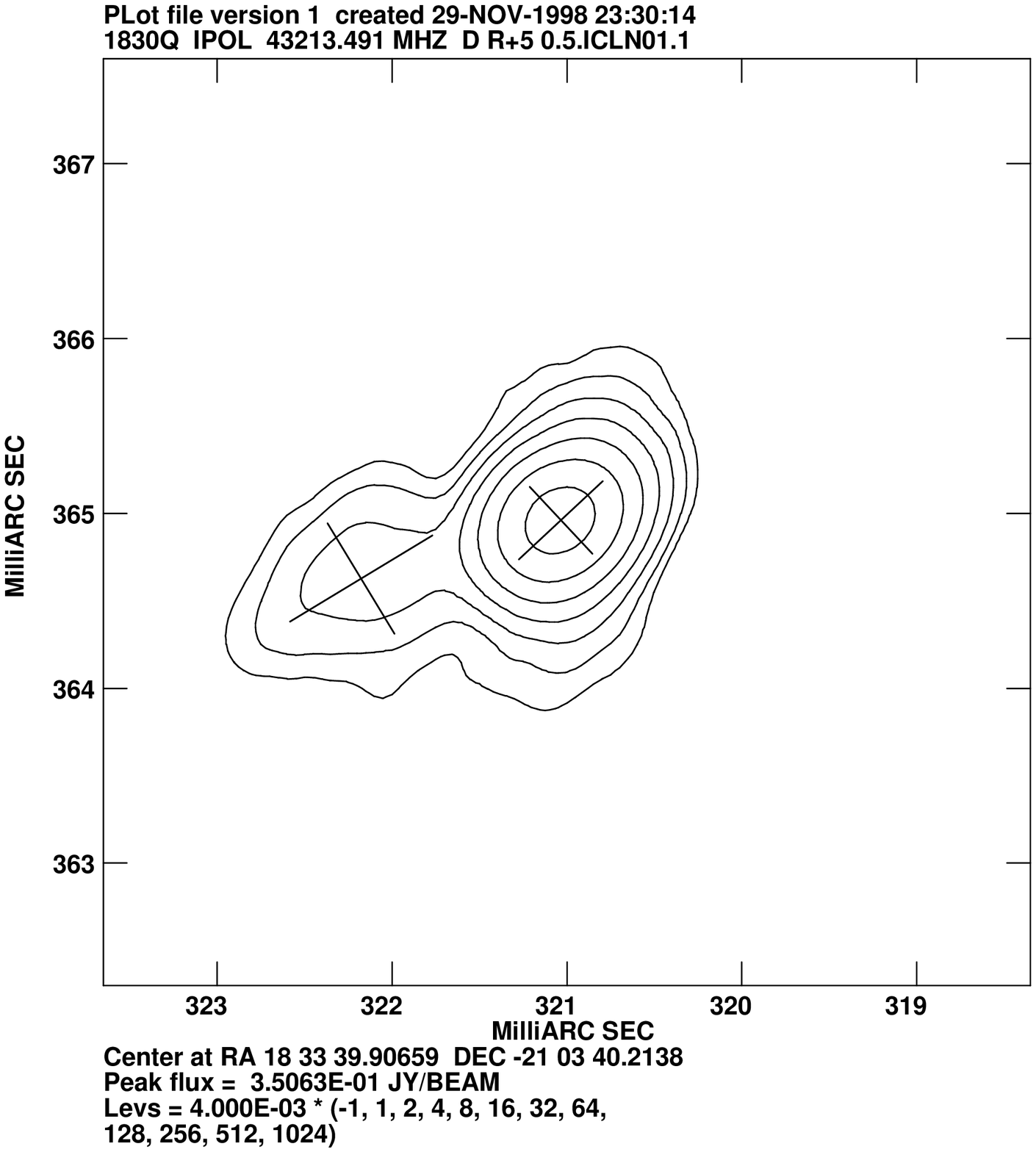}
\includegraphics[scale=0.13,angle=0]{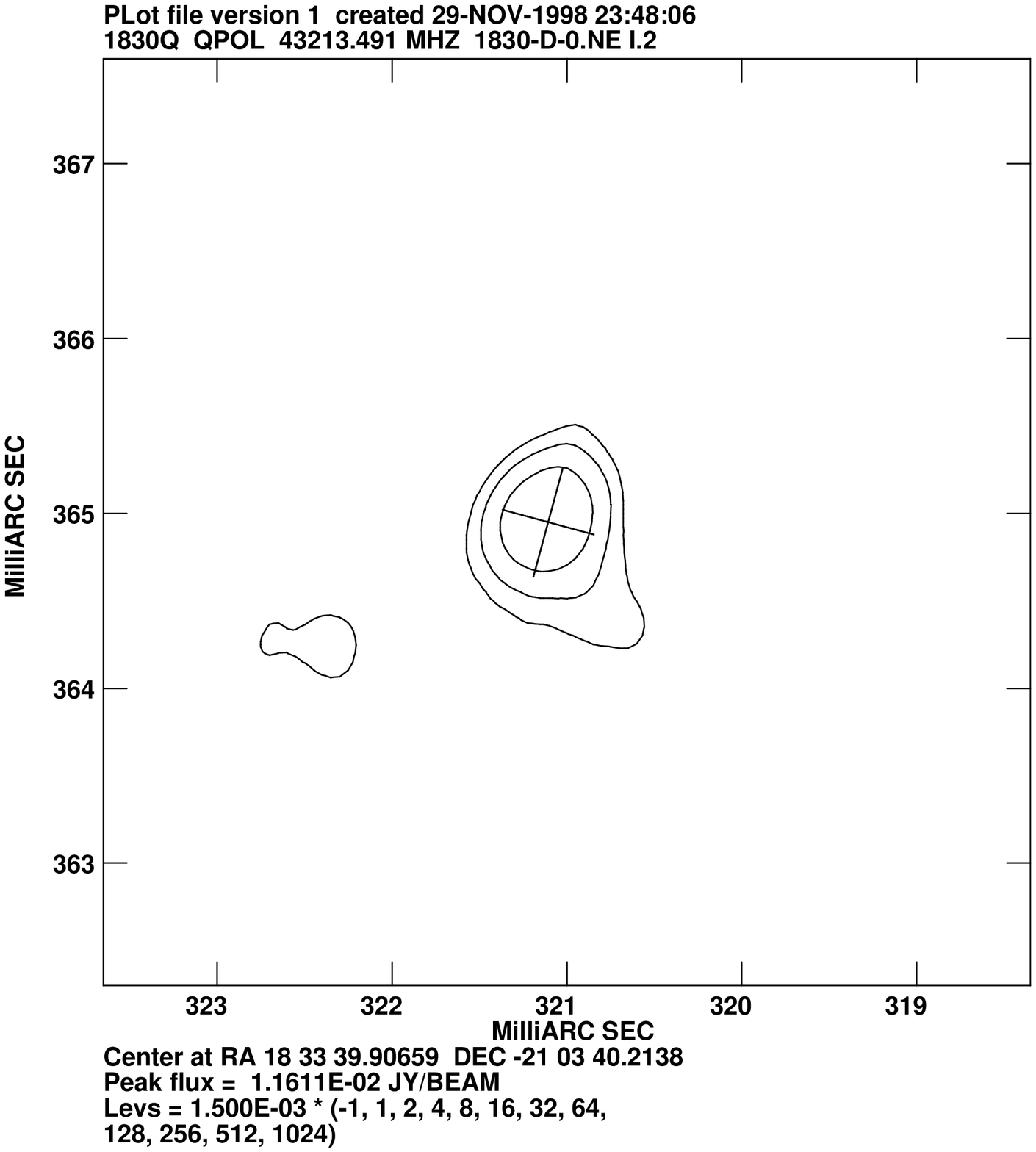}
\includegraphics[scale=0.13,angle=0]{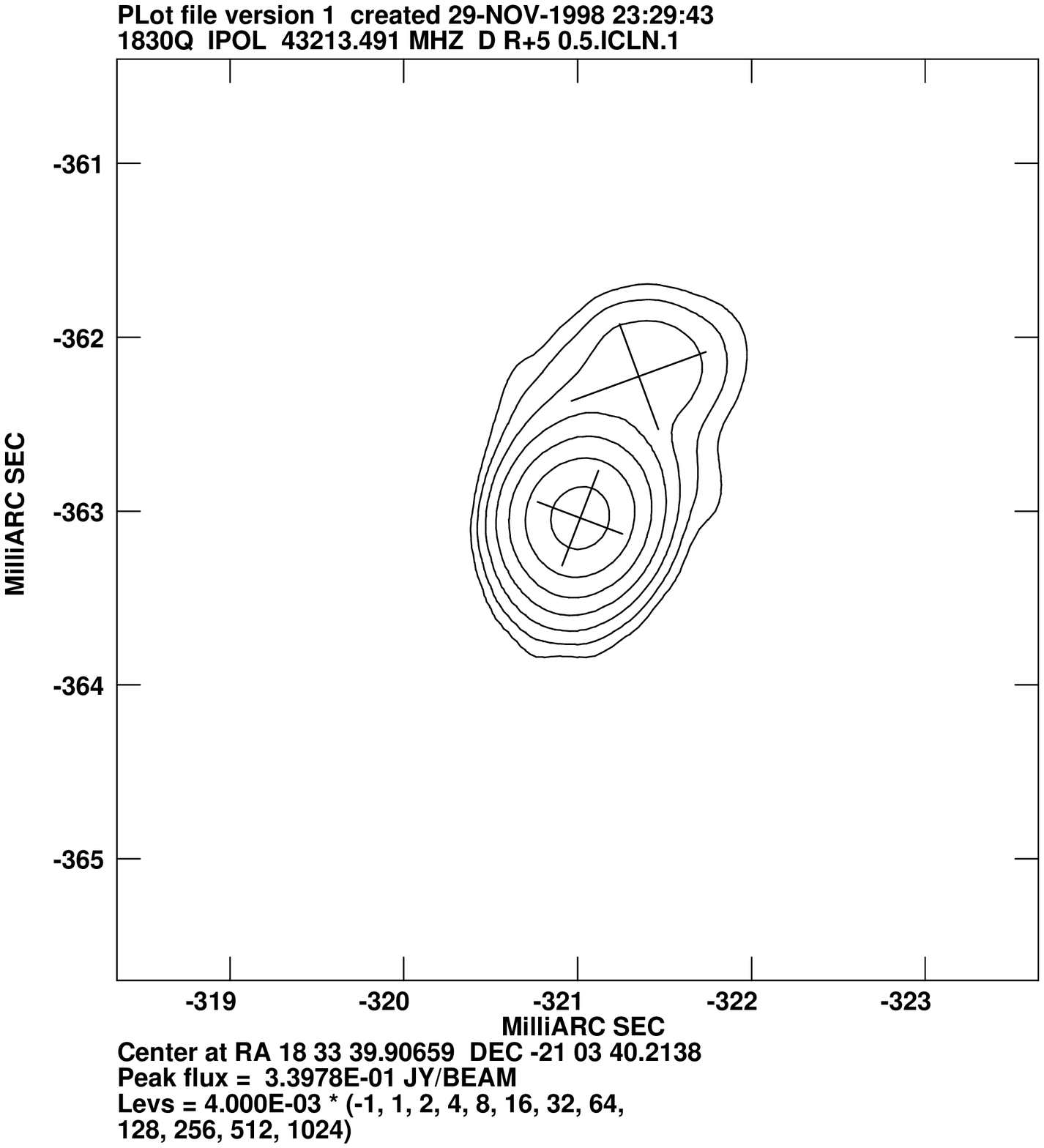}
\includegraphics[scale=0.13,angle=0]{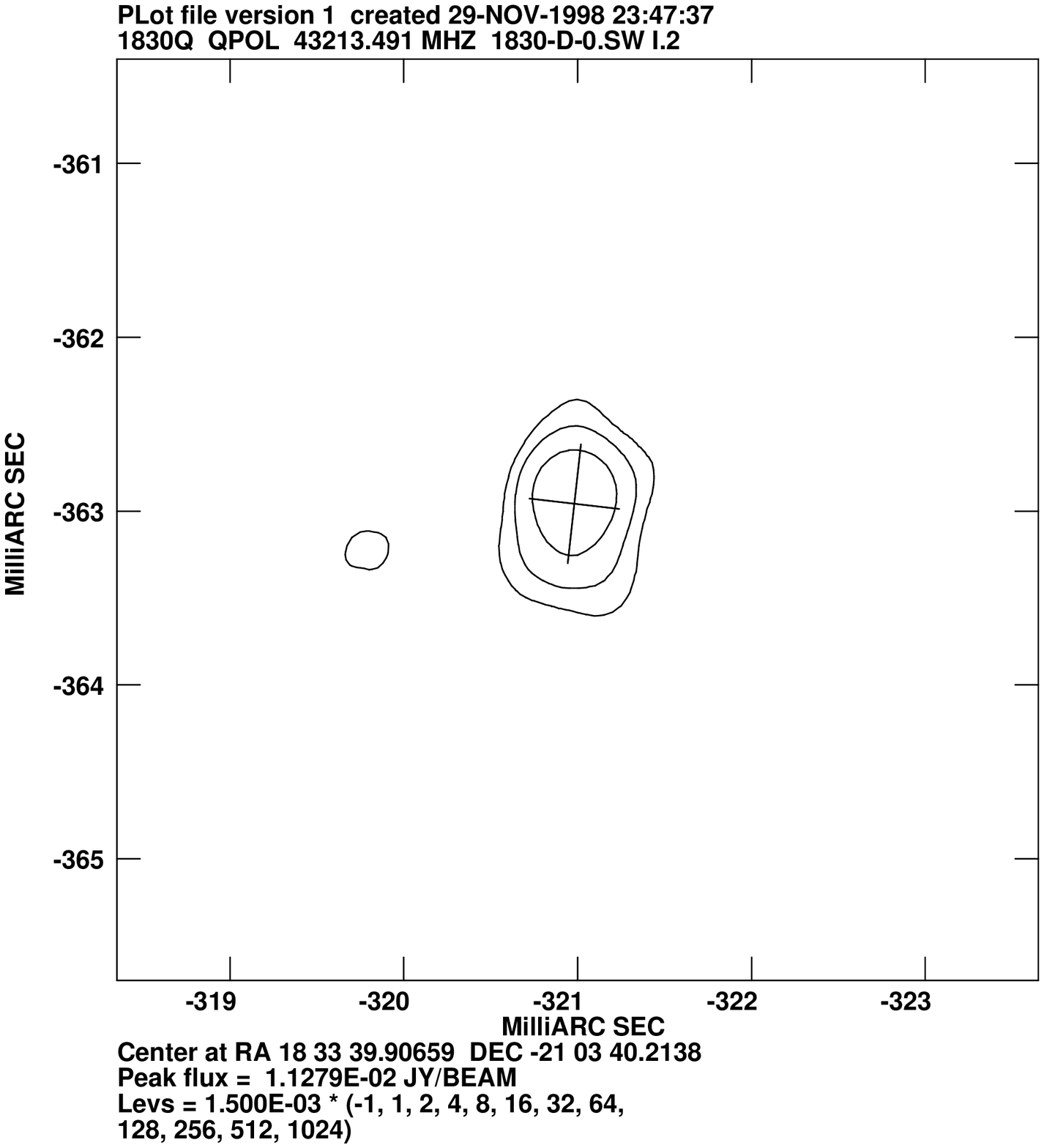}

\includegraphics[scale=0.13,angle=0]{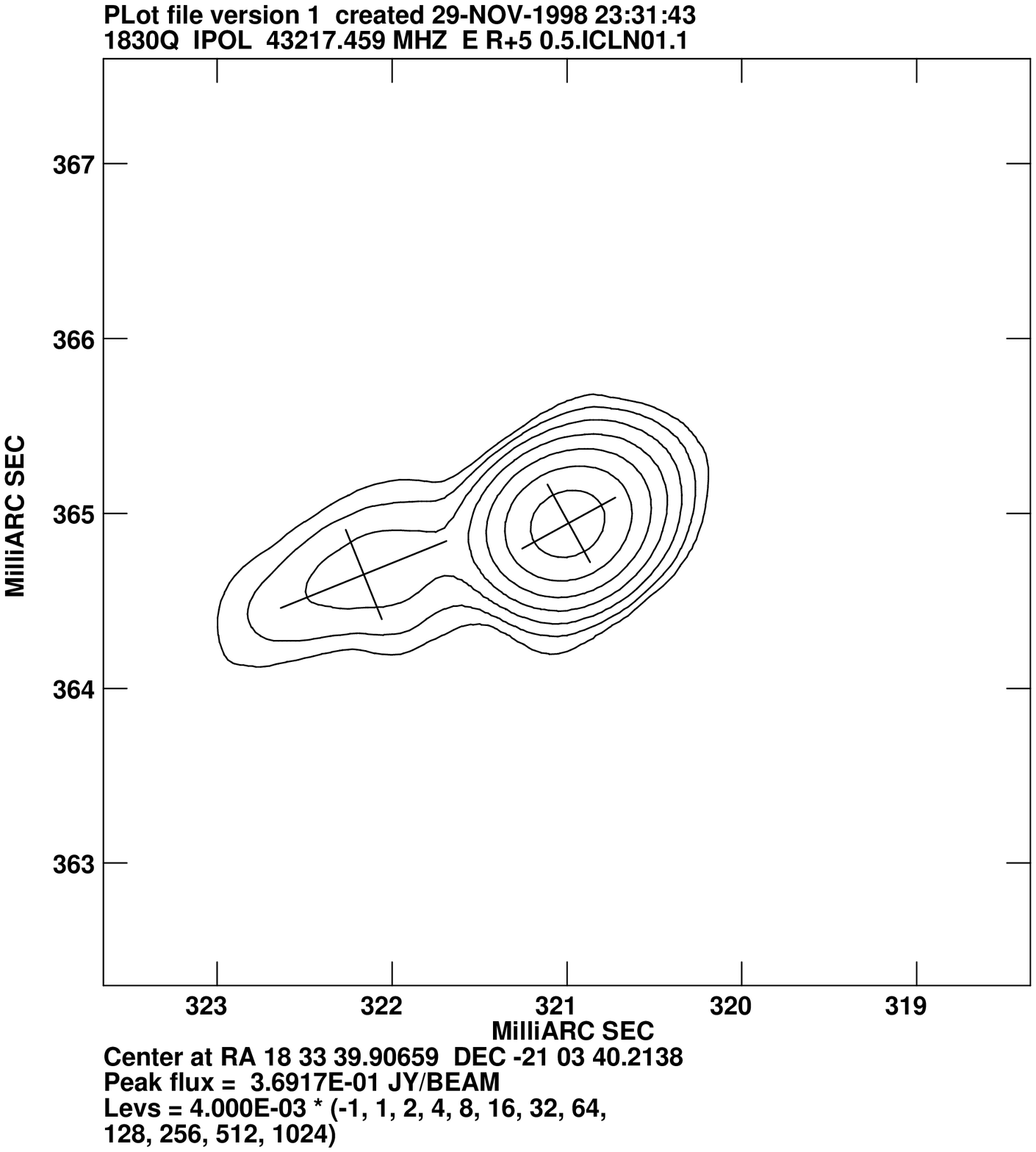}
\includegraphics[scale=0.13,angle=0]{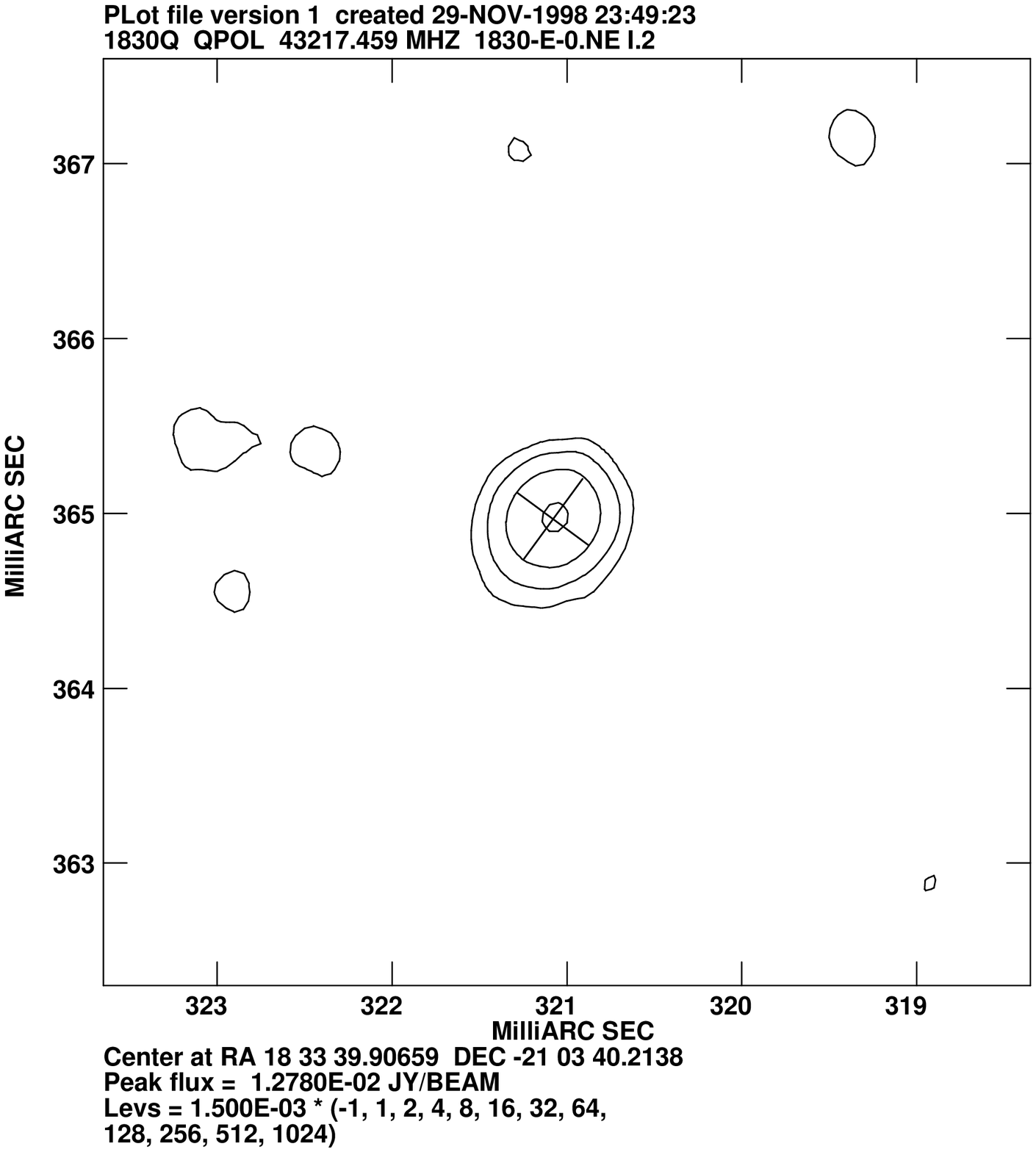}
\includegraphics[scale=0.13,angle=0]{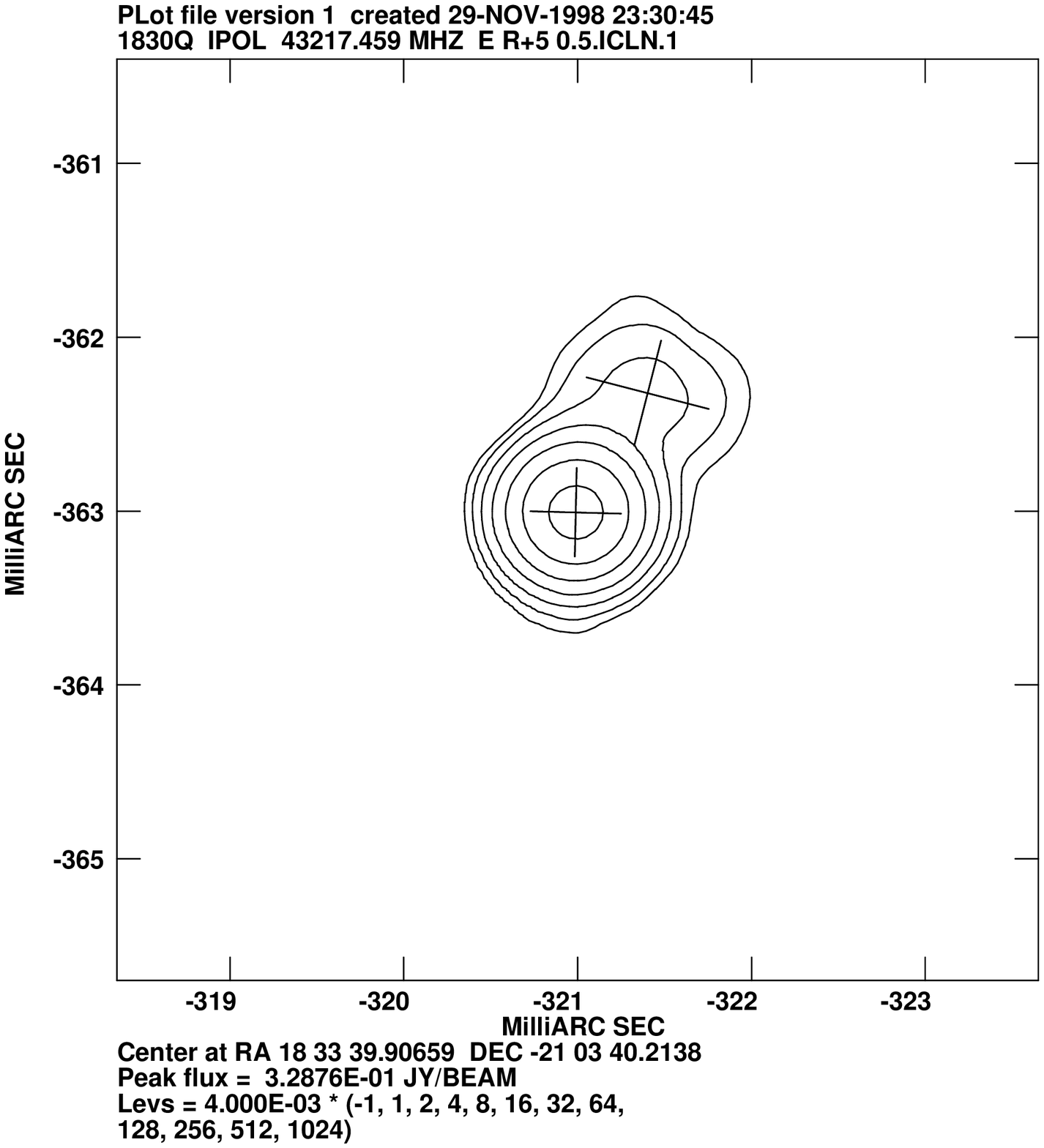}
\includegraphics[scale=0.13,angle=0]{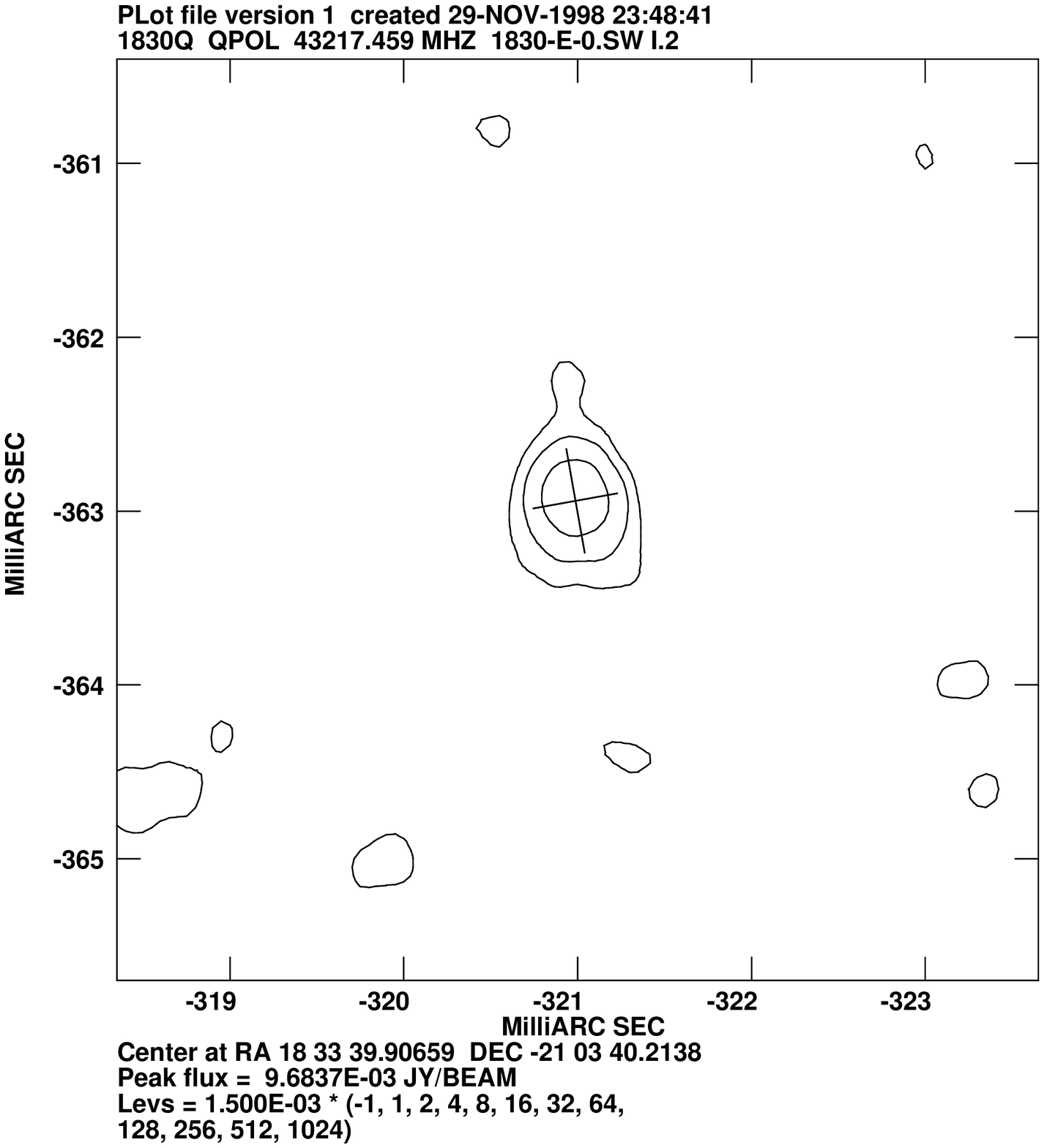}

\includegraphics[scale=0.13,angle=0]{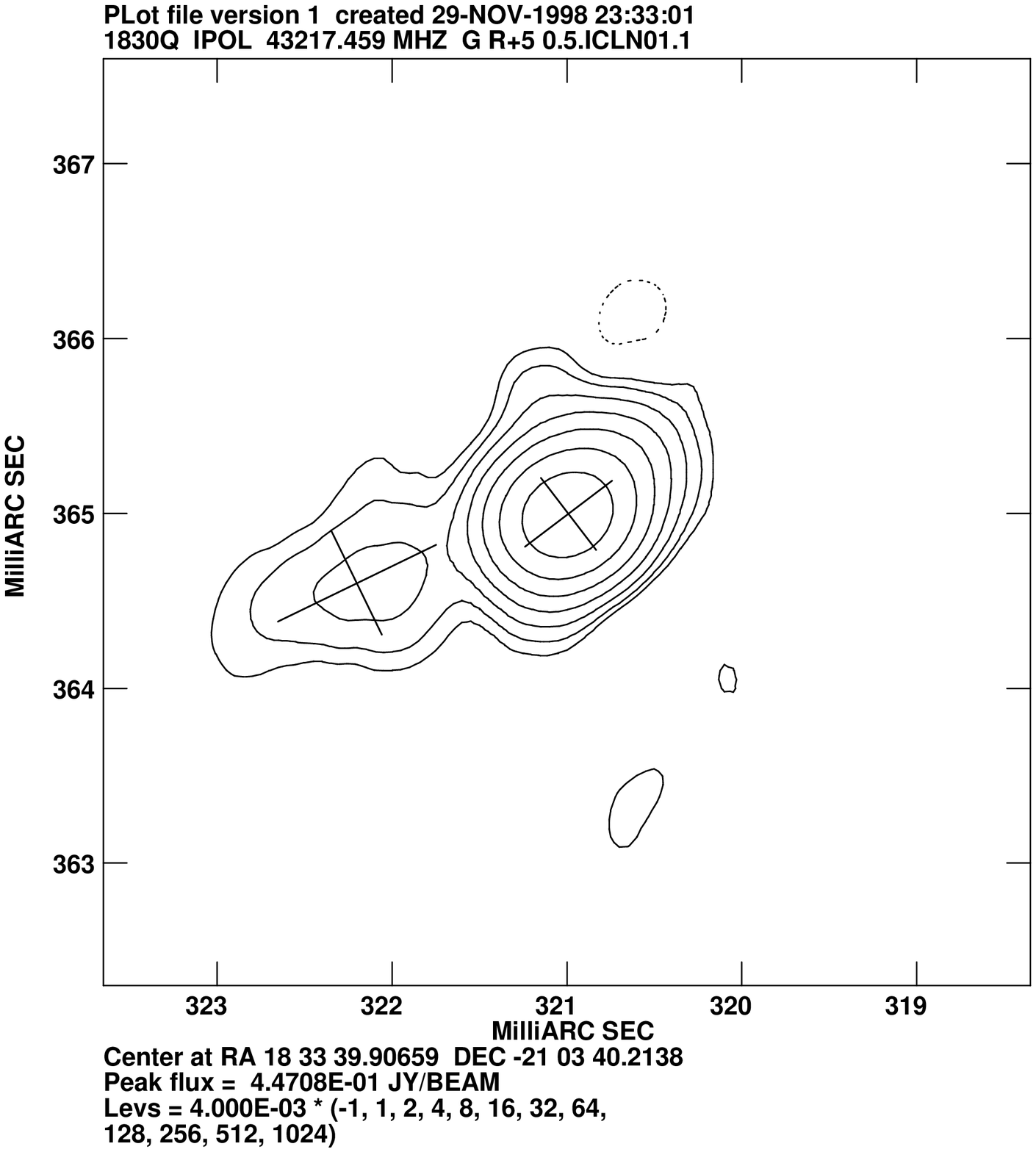}
\includegraphics[scale=0.13,angle=0]{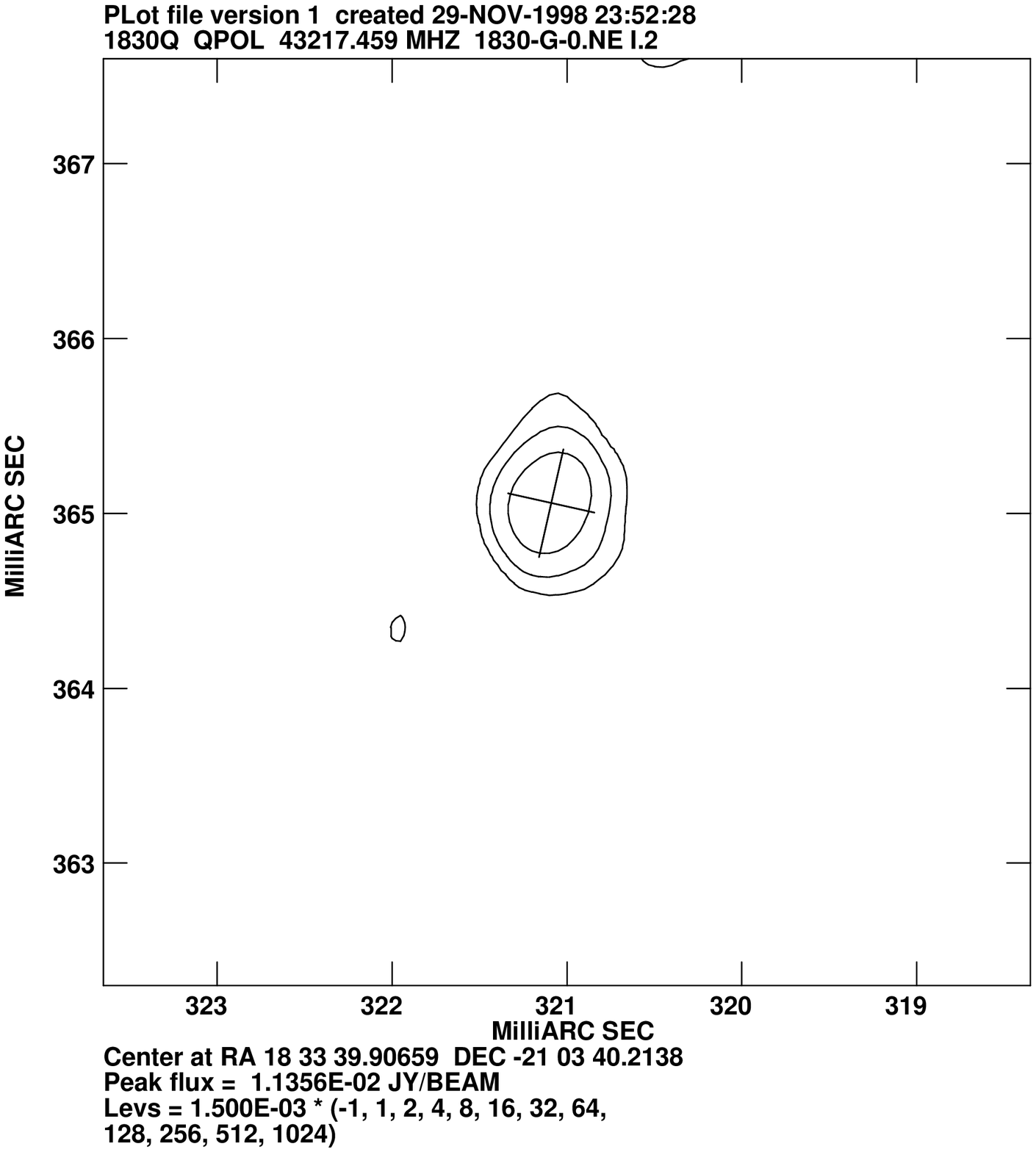}
\includegraphics[scale=0.13,angle=0]{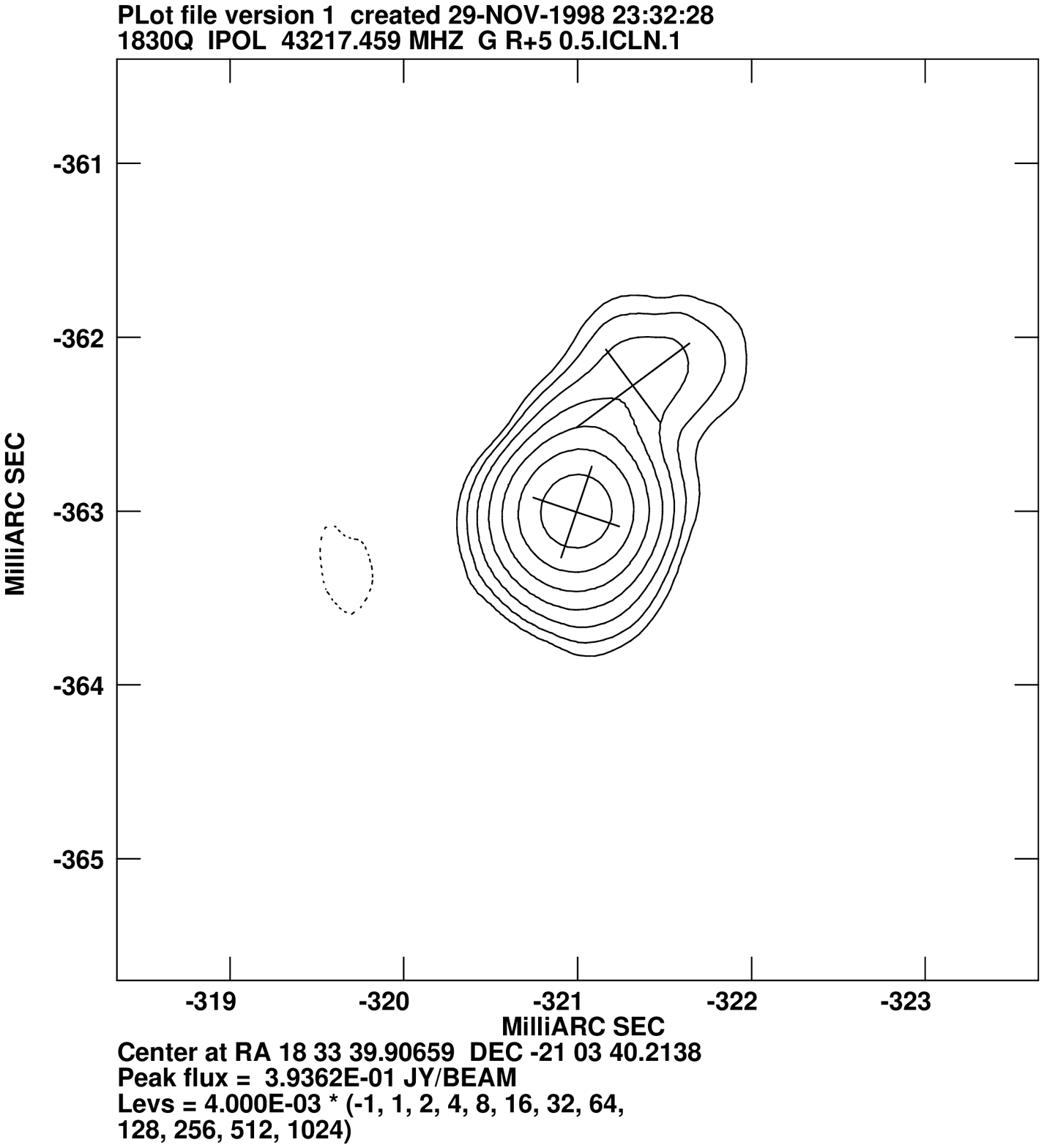}
\includegraphics[scale=0.13,angle=0]{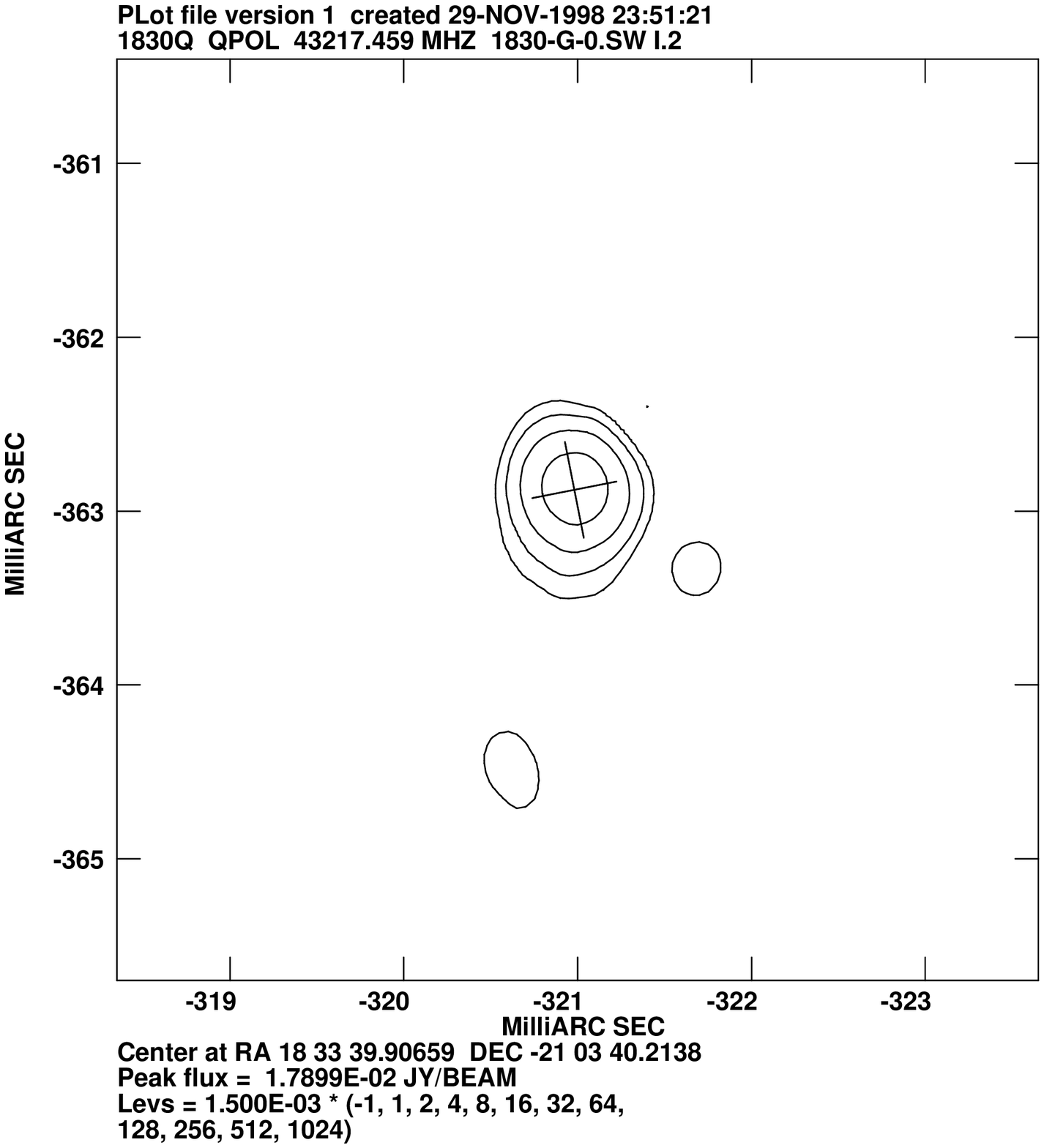}

\includegraphics[scale=0.13,angle=0]{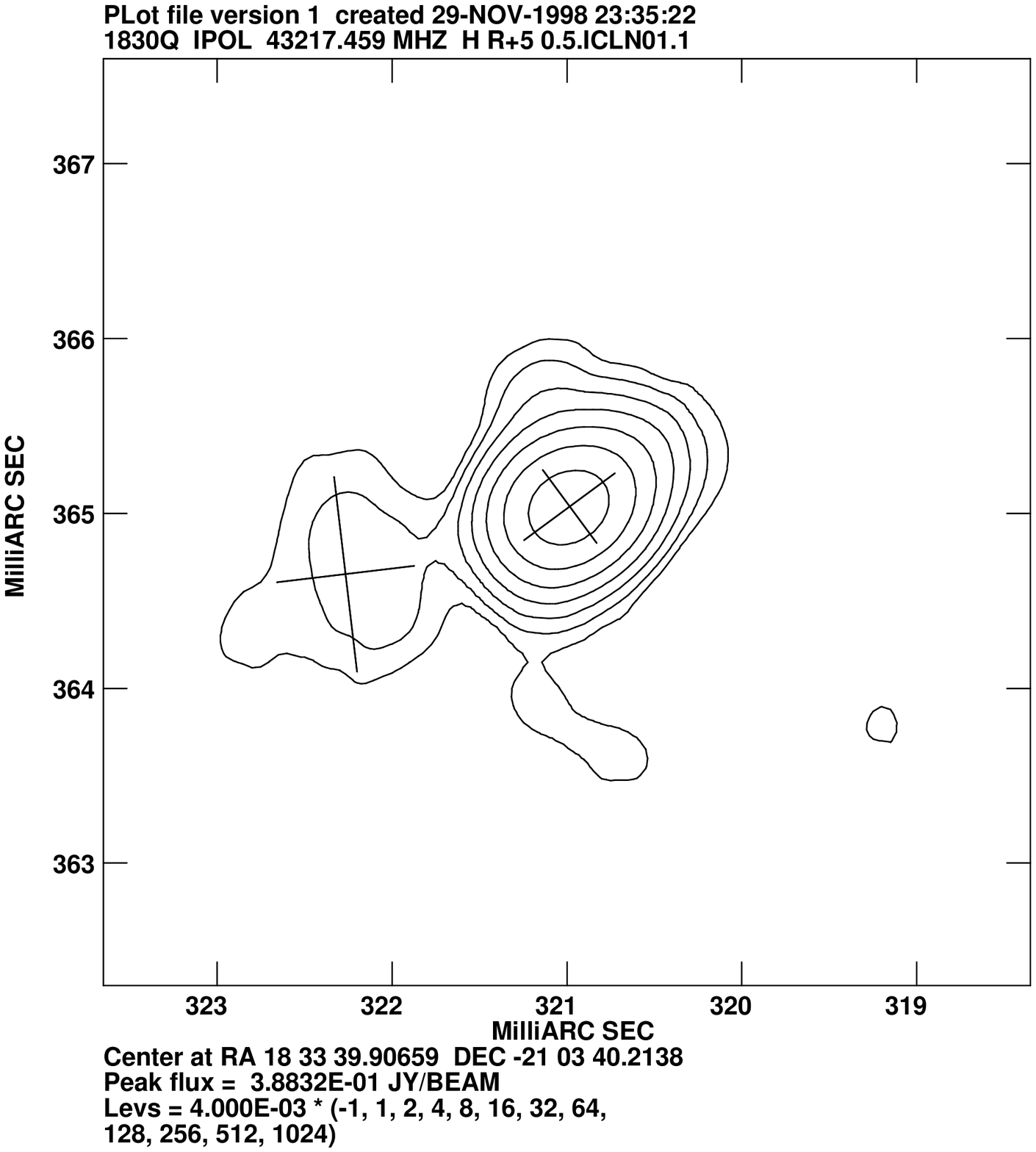}
\includegraphics[scale=0.13,angle=0]{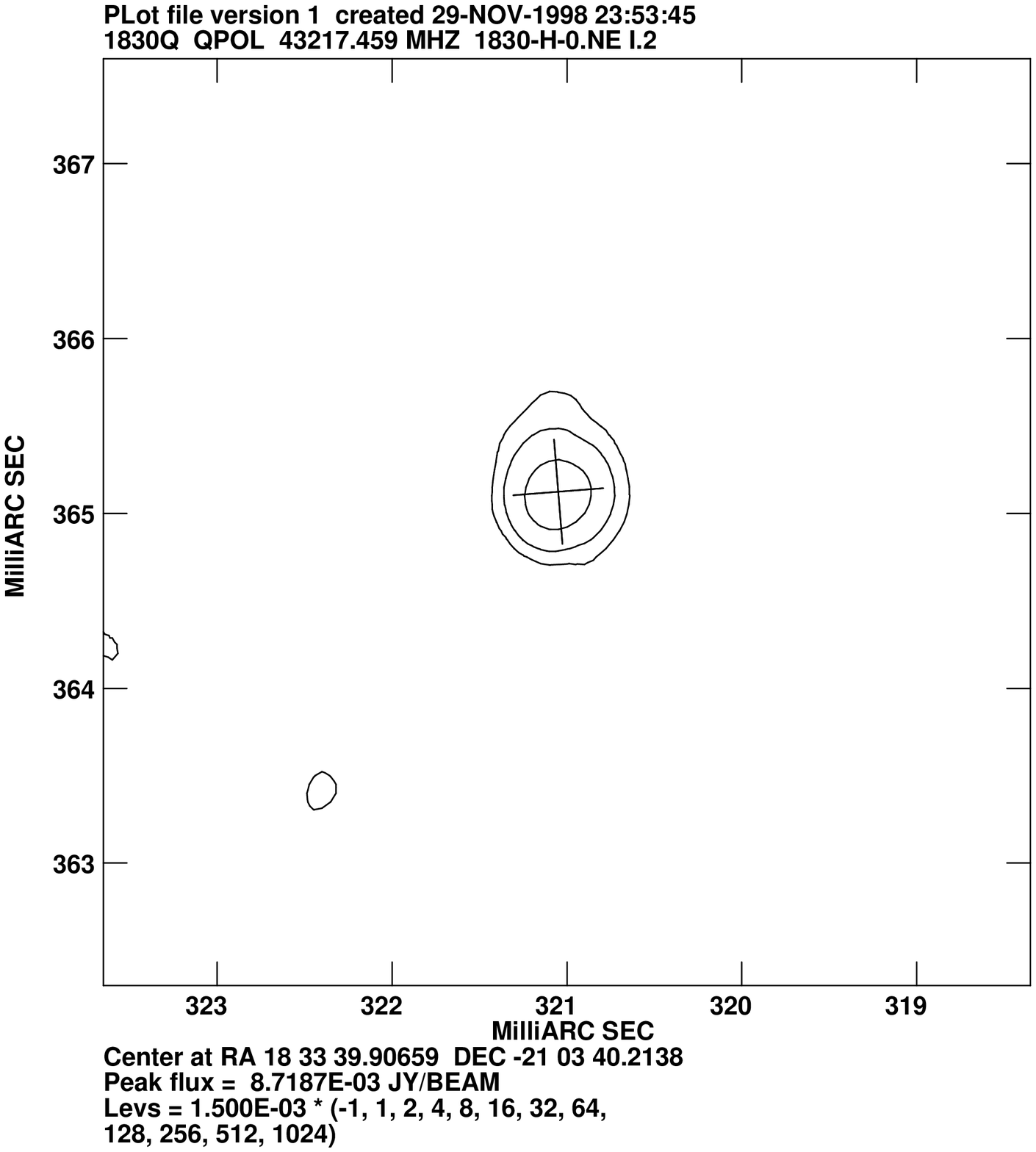}
\includegraphics[scale=0.13,angle=0]{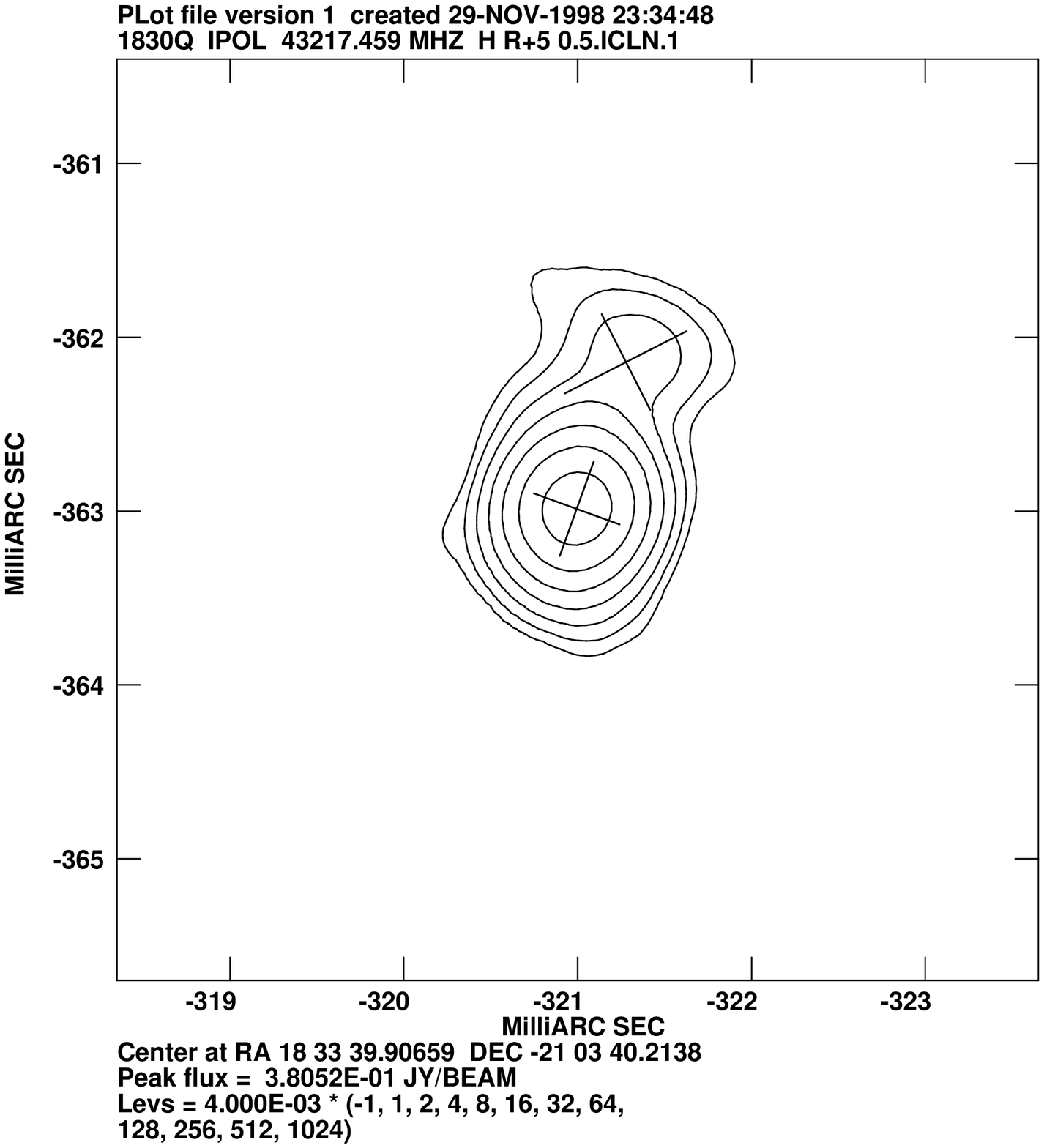}
\includegraphics[scale=0.13,angle=0]{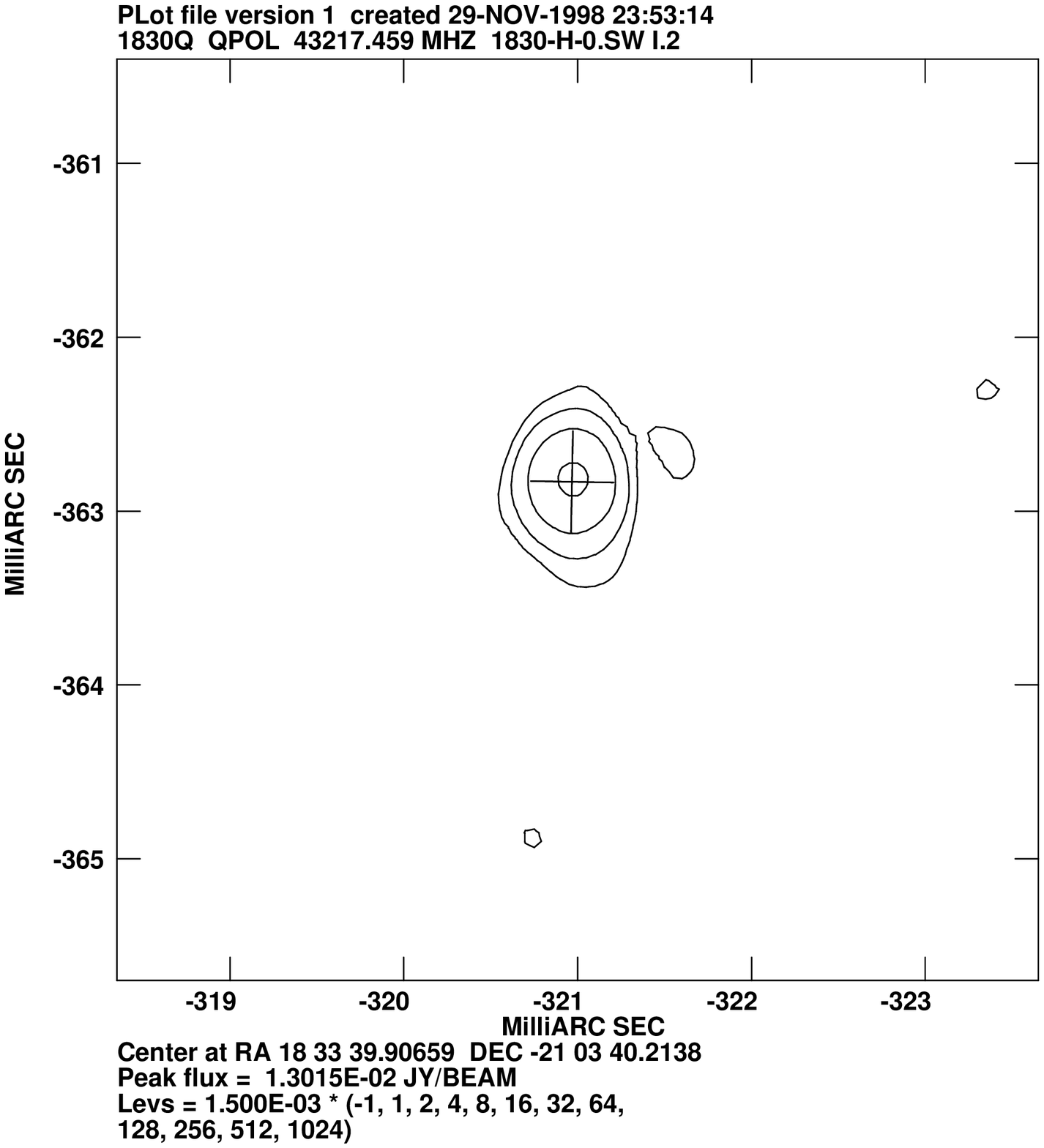}

\caption{From left to right we present:
(i) total intensity maps of the NE image, (ii) polarised intensity maps
of the NE image. (iii)  total intensity maps of the SW image and (iv)
polarised intensity maps of the SW image. The epoch of observation
increases from top to bottom. All the maps are naturally weighted
and the FWHM of the circular restoring beam is 0.5~mas.
Contours are spaced by factors of two in brightness, with the
lowest at three times the rms noise 4~mJy per beam (for the total
intensity maps) and 1.5 mJy per beam (for the polarised intensity
maps).}
\label{fig1}
\end{figure}

Contour maps of both PKS~1830-211~NE and SW, in total and
polarised intensity, for each epoch except the sixth (1997 April 03)
are shown in Fig.~\ref{fig1}. 
Superimposed on
these maps are the positions and sizes of the Gaussian fits as
determined by the AIPS task {\sc IMFIT}. The size of the crosses
represent the major and minor axes of the Gaussian components. 

Our wide-field approach to the data analysis permits us to produce maps
of both lensed images simultaneously, thus allowing us to measure with
high precision the angular separation between the central peaks in the radio
images. 

\begin{figure}[htbp]
\centering
\includegraphics[scale=0.6,angle=-90]{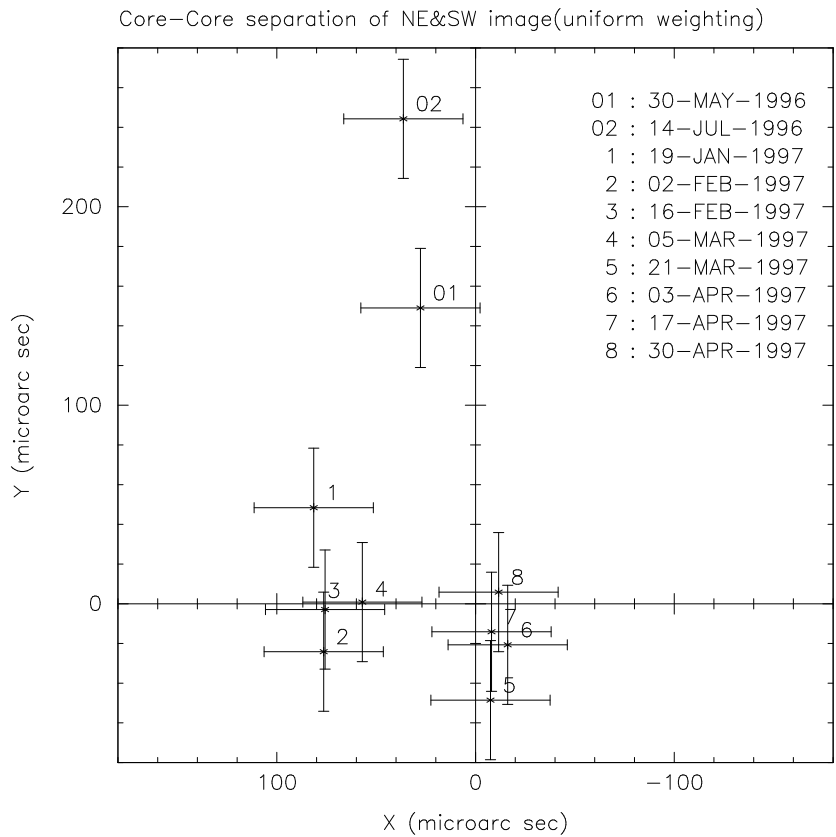}
\caption{The position of the total intensity peak in the NE image relative 
to the total intensity peak in the SW image. 
 The origin on the plot is at a nominal
position (642000,728000) in microarcsec with respect to SW core, which is 
the separation of the two sub-fields of the NE and SW images.} 
\label{fig2}
\end{figure}

We have measured the angular separation of the NE and SW image (NE--SW)
by fitting Gaussian components to our highest-resolution, uniformly-weighted,
total-intensity maps. In Fig.~\ref{fig2}
we present the
position of the peak in the NE image relative to the peak in the SW
image.We have also included the results of previous 7\,mm VLBI observations
made by \citet{Garrett97} in 1996.

We have estimated the errors on all these separation measurements by
comparing measurements from sub-sets of the data for a given epoch.
These suggest the separation measurements are accurate to 1/10 of the
major axis of the uniformly weighted fitted beam i.e. $\sim
30\,\mu$as. 

What are the possible explanations for this change in the image
separation ? 

First let us consider effects that are intrinsic to the background
source. (In this case changes will appear in both images, separated by
the time delay, and the angular separation changes result from a
combination of both). In this context it is useful to note that for a
simple FRW universe ($q_{0}=0$, $\Lambda=0$, $H_{0}=65$ km\,s$^{-1}$\,Mpc$^{-1}$), 
a shift of $\sim 80\,\mu$as (the largest shift measured between
epochs separated by $\sim 2$~weeks) corresponds to a linear distance
of $\sim 0.8$~pc at $z_{s}=2.507$. If one assumes that the lens
provides a  magnification factor of $\sim 10$
\citep{Kochanek92, Nair93}, the linear distance then scales to $\sim
0.08$~pc. Given PKS~1830-211's blazar characteristics, the most obvious
interpretation is that the centroid of the peak in the brightness
distribution of the core is changing on relatively short time-scales,
perhaps for example, as shock fronts propagate along a continuous jet,
appearing as bright and (later) fading regions of radio emission.  A
less conventional scenario is that it is the mm-VLBI ``core'' (i.e.
the base of the jet) itself that is moving, as the effectiveness of
the collimation mechanism changes.  Since this area of jet physics
remains poorly understood (see \citet{Marscher95} for a recent
review), and since measurements with this sort of linear resolution
have not previously been possible in this type of source, it is not
clear to us whether movements of the order of $\sim 0.08$~pc are
reasonable or not.

Extrinsic explanations might include the effect of scattering by
ionised gas encountered along the line of sight, specifically in
the lens galaxy or our own galaxy. Since PKS~1830-211 lies close to the
galactic plane, the effect of the ISM in our own galaxy is expected to
dominate \citep{Walker96}. \citet{Jones96} have measured the
interstellar broadening of the SW image between 18 and 1.3~cm and
estimate that the SW deconvolved core size is proportional to
$\lambda^{\sim 2}$, as predicted by Interstellar Scattering (ISS) theory.  At
1.3~cm they measure a size of the SW image of $\sim 0.6 \times 0.2
$~mas. In our observations the size of the SW (and the NE) image change
with epoch. However, a typical value for the core size is $\sim 0.26
\times 0.13$ which is much larger than we might expect 
assuming a $\lambda^{2}$ relation. Indeed the scaling in size between
1.3~cm and 0.7~mm is almost
linear with $\lambda$, as expected from simple models of synchrotron
radio emission. Hence, we suspect that the measured source size is
dominated by its internal radio structure, rather than scattering effects.

However, in addition to image broadening, the effect of ISS can, in
principle, also produce ``image wander'' --- an apparent shift in the
position of a source \citep{Rickett90}. This effect is considered to be
small: an order of magnitude smaller than the scattering size itself.
Indeed, observations of compact sources lying close to the galactic
centre (where the effects of ISS should be severe) bear this out: for example,
$\lambda 18$~cm VLBI observations of masers located close to the
galactic centre \citep{Gwinn88} show that for water masers the r.m.s.
wander of individual spots is $< 18\,\mu$as over the course of 6
months. Since the effect of image-wander would also scale as $\lambda^{\sim2}$,
we suspect that conventional ISS is not a compelling explanation for
the changes in image separation that we measure.  We also observe
changes in the image separation with respect to the peaks in polarised
intensity. These changes are less reliable than those observed in total
intensity (since the polarised flux is very much fainter) but the
initial indications are that the changes in the image separation in
total intensity and polarised intensity are unrelated.

Milli-lensing produced by massive ($10^{3}$--$10^{4}\,M_{\odot}$) compact
objects in the halo of the lens can certainly introduce shifts of $\sim
80\,\mu$as, but changes in the separation would be measured on
relatively long time-scales: hundreds of years rather than the weeks or
months observed here. 

Another possibility is that the transverse velocity of the lens galaxy 
across the sky could introduce a relative proper motion between the 
NE and SW images. For highly magnified four-image lens systems the 
proper motions are expected to be a $\sim$~few tens~$\mu$as\,yr$^{-1}$
\citep{Kochanek96}, but for two-image systems such as PKS~1830-211
the motion is expected to be an order of magnitude smaller. 

In summary, the changes in the measured image separation are most likely
due to changes in the brightness distribution of the background radio
source. The detection of source evolution on these short time-scales
would be impossible if it were not for the fact that this is a lensed
system which provides us with a magnified view and closely spaced
multiple images that allow accurate relative position measurements to
be made. If, as we strongly suspect, the changes we observe are due to
internal motions in the radio structure on scales of $> 0.08$~pc in
$\sim 2$~weeks, this implies (unlensed) 
superluminal velocities in the rest frame
of the background source of $> 3c$.

\end{document}